\documentclass[12pt]{article}
\usepackage{amsmath,amssymb,mathrsfs,cite,color}
\usepackage{graphicx} 
\usepackage{float}
\newcommand{\tr}{\operatorname{tr}}

\newcommand{\adj}{\operatorname{adj}}

\newcommand{\mycomment}[1]{}

\allowdisplaybreaks
\numberwithin{equation}{section}

\begin{document}
\begin{titlepage}
\renewcommand{\thefootnote}{\fnsymbol{footnote}}

  \begin{flushright}
  UTHEP-778\\
  KEK-TH-2500
\end{flushright}
  
\begin{center}
{\large \bf
  The dynamics of zero modes in lattice gauge theory\\
  --- difference between ${\rm SU}(2)$ and ${\rm SU}(3)$ in 4D
}
\end{center}
\vspace{3mm}

\begin{center}
         Yuhma A{\sc sano}$^{1,2)}$\footnote
         { E-mail address : asano@het.ph.tsukuba.ac.jp}
         and
         Jun N{\sc ishimura}$^{3,4)}$\footnote
          { E-mail address : jnishi@post.kek.jp}

          \par          \vspace{7mm}
          
$^{1)}${\it
Faculty of Pure and Applied Sciences, University of Tsukuba,\\
1-1-1 Tennodai, Tsukuba, Ibaraki 305-8571, Japan
}

$^{2)}${\it
Tomonaga Center for the History of the Universe, University of Tsukuba,\\
1-1-1 Tennodai, Tsukuba, Ibaraki 305-8571, Japan
}



          
$^{3)}$\textit{Theory Center,
Institute of Particle and Nuclear Studies,}\\
{\it High Energy Accelerator Research Organization (KEK),\\
1-1 Oho, Tsukuba, Ibaraki 305-0801, Japan} 


$^{4)}$\textit{Department of Particle and Nuclear Physics,}\\
\textit{School of High Energy Accelerator Science,}\\
{\it Graduate University for Advanced Studies (SOKENDAI),\\
1-1 Oho, Tsukuba, Ibaraki 305-0801, Japan} 
\end{center}

\vspace{0.5cm}

\begin{abstract}\noindent
The dynamics of zero modes in gauge theory
is highly nontrivial due to its 
nonperturbative nature even in the case where 
the other modes can be treated perturbatively.
One of the related issues concerns the possible instability of 
the trivial vacuum $A_\mu(x)=0$ due to the existence of 
nontrivial degenerate vacua known as ``torons''.
%
%
Here we investigate this issue for the 4D SU(2) and SU(3) 
pure Yang-Mills theories on the lattice
by explicit Monte Carlo calculation of the Wilson loops 
and the Polyakov line
at large $\beta$.
While we confirm the leading $1/\beta$
predictions obtained around the trivial vacuum
in both SU(2) and SU(3) cases,
we find that the subleading term
vanishes only logarithmically in the SU(2) case
unlike the power-law decay in the SU(3) case.
%
In fact, the 4D SU(2) case is marginal
according to the criterion by Coste et al.
Here we show that the trivial vacuum dominates in this case
due to large fluctuations of the zero modes around it,
thereby providing a clear understanding of the observed behaviors.
\end{abstract}
\setcounter{footnote}{0}
\end{titlepage}

\section{Introduction}

Zero modes in gauge theory play an important role
when one considers the theory in a small box
with periodic boundary conditions.
Their dynamics is highly nontrivial, however, due to its 
nonperturbative nature even in the case where 
the other modes can be treated perturbatively.
The importance of zero modes can be
most clearly seen by recalling the 
large-$N$ reduced model
\cite{Eguchi:1982nm}, which implies that 
the zero-mode effective theory can be actually equivalent to the 
original gauge theory
under certain 
conditions \cite{Gonzalez-Arroyo:1982hyq,Gonzalez-Arroyo:2010omx}.
Furthermore, the maximally supersymmetric version of the large-$N$ 
reduced model \cite{Ishibashi:1996xs}
is conjectured to be a nonperturbative formulation of superstring theory,
in which (3+1)-dimensional expanding space-time is expected to 
emerge \cite{Anagnostopoulos:2022dak}.


One of the issues related to the dynamics of zero modes
in lattice gauge theory
concerns 
the existence of 
nontrivial degenerate vacua known as ``torons'',
which have constant diagonal link variables 
in each direction up to a gauge transformation.
All these configurations minimize the plaquette action,
which implies that they represent the orbit of vacua
or the moduli space of the gauge field configurations.
At weak coupling, 
the perturbative expansion can be performed 
by integrating out all the fluctuations
including zero modes
around each toron configuration,
and one is left with the integration over the toron configurations.
Then an important question is
whether the trivial vacuum 
$U_{n,\mu}={\bf 1}$
dominates
over the other toron configurations or not.

This question is answered in the affirmative
for 4D ${\rm SU}(3)$ lattice gauge theory \cite{Coste:1985mn},
and the leading $1/\beta$ correction to the Wilson loop have been
obtained by perturbative expansion around the trivial vacuum.
The 4D ${\rm SU}(2)$ case, on the other hand, 
turned out to be marginal
according to the same criterion.


In this paper, we first show our Monte Carlo results
for the Wilson loops at large $\beta$ in the 4D SU(2) and SU(3) cases
and compare them against the O$(1/\beta)$ predictions
obtained by perturbation theory around the trivial vacuum.
In the ${\rm SU}(3)$ case,
our results 
approach the perturbative predictions
as $\beta \rightarrow \infty$
with the deviation being 
suppressed
by $\beta^{-1/2}$ compared to the leading O($1/\beta$) correction.
In the ${\rm SU}(2)$ case, on the other hand,
our results approach the perturbative predictions as well,
but the approach turns out to be 
much slower than in the ${\rm SU}(3)$ case.
The deviation is 
suppressed only by O($1/\log \beta$)
compared to the leading O($1/\beta$) correction.
This suggests that the trivial vacuum dominates
as $\beta \rightarrow \infty$ in the SU(2) case as well
although the fluctuations around it vanish
much more slowly with increasing $\beta$ 
than in the ${\rm SU}(3)$ case.

Here
we attempt to 
understand
this behavior of the 4D ${\rm SU}(2)$
lattice gauge theory
by the effective theory for the zero modes
around the trivial vacuum,
which takes 
the form
of the large-$N$ reduced model at the leading order.
In fact, it was noticed in Section 3.3 of Ref.~\cite{Coste:1985mn} that
the partition function of 
the zero-mode effective theory
%
has logarithmic divergence
as $\beta\to\infty$
in the 4D ${\rm SU}(2)$ case\footnote{In Ref.~\cite{Fodor:2012td},
non-perturbative treatment of the zero modes has been discussed 
in the context of the gradient flow 
used in a running coupling constant scheme.
In particular, the logarithmic behavior in the SU(2) case has
been discussed by using a simple example \cite{Fodor:2012qh},
in which $1/\log \beta$ appears in the perturbative expansion
of the exact result.}.
This is actually due to the 
power-law tail (with the power of $-1$)
of the eigenvalue distribution \cite{KS99Ei}
in the reduced model,
which was investigated in the context of matrix models for superstrings.
We argue that this power-law tail 
is responsible for
not only the dominance of
  the trivial vacuum
  as $\beta \rightarrow \infty$ but also the unusual
  finite $\beta$ effects.

  We also calculate the Polyakov line at large $\beta$ 
in order to confirm our conclusion.
In particular, we find that the distribution of the Polyakov line
is peaked near unity in both SU(2) and SU(3) cases, but
the difference is that the distribution in the SU(2) case has a long tail,
which vanishes only very slowly at large $\beta$.

The rest of this paper is organized as follows.
In section \ref{sec:wilson-loop},
we present our Monte Carlo results for the Wilson loops at large $\beta$,
and compare them with perturbative predictions.
In section \ref{sec:eff-theory},
we review the effective theory for the zero modes,
which explains our results in the SU(3) case. 
In section \ref{sec:reduced-YM},
we discuss how 
our results in the SU(2) case
can be understood
from the viewpoint of the zero-mode effective theory.
In section \ref{sec:polyakov}, we 
show
our results for the Polyakov line,
which confirm our understanding.
Section \ref{sec:summary} is
devoted to a summary and discussions.

\section{Wilson loops v.s.\ perturbative predictions}
\label{sec:wilson-loop}

In this section, we present our Monte Carlo results for the Wilson loops
and compare them with perturbative predictions at large $\beta$.
Let us consider an
${\rm SU}(N)$ gauge theory
on an $L_1\times \cdots \times L_D$ lattice
with periodic boundary conditions with the action
\begin{align}
 S_{\rm g}
 =-\frac{\beta}{2N}\sum_n\sum_{\mu,\nu\ne\mu}
 \tr[U_{n,\mu}U_{n+\hat\mu,\nu}U_{n+\hat\nu,\mu}^{\dag}U_{n,\nu}^{\dag}-1] \  ,
 \label{plaq-action}
\end{align}
where $\beta$ is the inverse coupling constant squared.
The Wilson loop of size $R_1\times R_2$
is calculated by perturbation theory around the trivial vacuum
$U_{n,\mu}={\bf 1}$
neglecting higher order terms than O($\beta^{-1}$) 
as \cite{Coste:1985mn}
\begin{equation}
 W(R_1,R_2)
 =1-\frac{c_1^{\rm (tot)}}{\beta}
 + {\rm O}(\beta^{-\frac{3}{2}}) \  .
 \label{WL-expansion}
\end{equation}
The coefficient
$c_1^{\rm (tot)} = c_1^{\rm (zero)} + c_1^{\rm (nonzero)}$
consists of
the zero-mode contribution
\begin{align}
 c_1^{\rm (zero)}
 &=  \frac{N^2 -1}{2(D-1)V} \, (R_1R_2)^2
 \label{c1-0}
\end{align}
and the non-zero-mode contribution
\begin{align}
 c_1 ^{\rm (nonzero)}
  = \frac{N^2-1}{8V} 
 \sum_{k\neq 0}
 \frac{1}{\sum_\mu \sin^2 \frac{k_\mu}{2}} 
 \left( \left| e^{ik_1R_1}-1 \right| ^2
 \left| \sum_{m=0}^{R_2-1}e^{ik_2m} \right| ^2
 + (1\leftrightarrow 2)
 \right) \ ,
 \label{c1-n}
\end{align}
where $k_\mu=\frac{2\pi}{L_\mu}n_\mu$
denotes the lattice momentum with $n_\mu$ being an integer.
Note that 
the nonzero-mode contribution vanishes
for a maximal Wilson loop with $R_1=L_1$ and $R_2=L_2$
since
$e^{ik_1R_1}=e^{ik_2R_2}=1$ in that case.

\begin{figure}[H]
 \centering
 \includegraphics[scale=0.45]{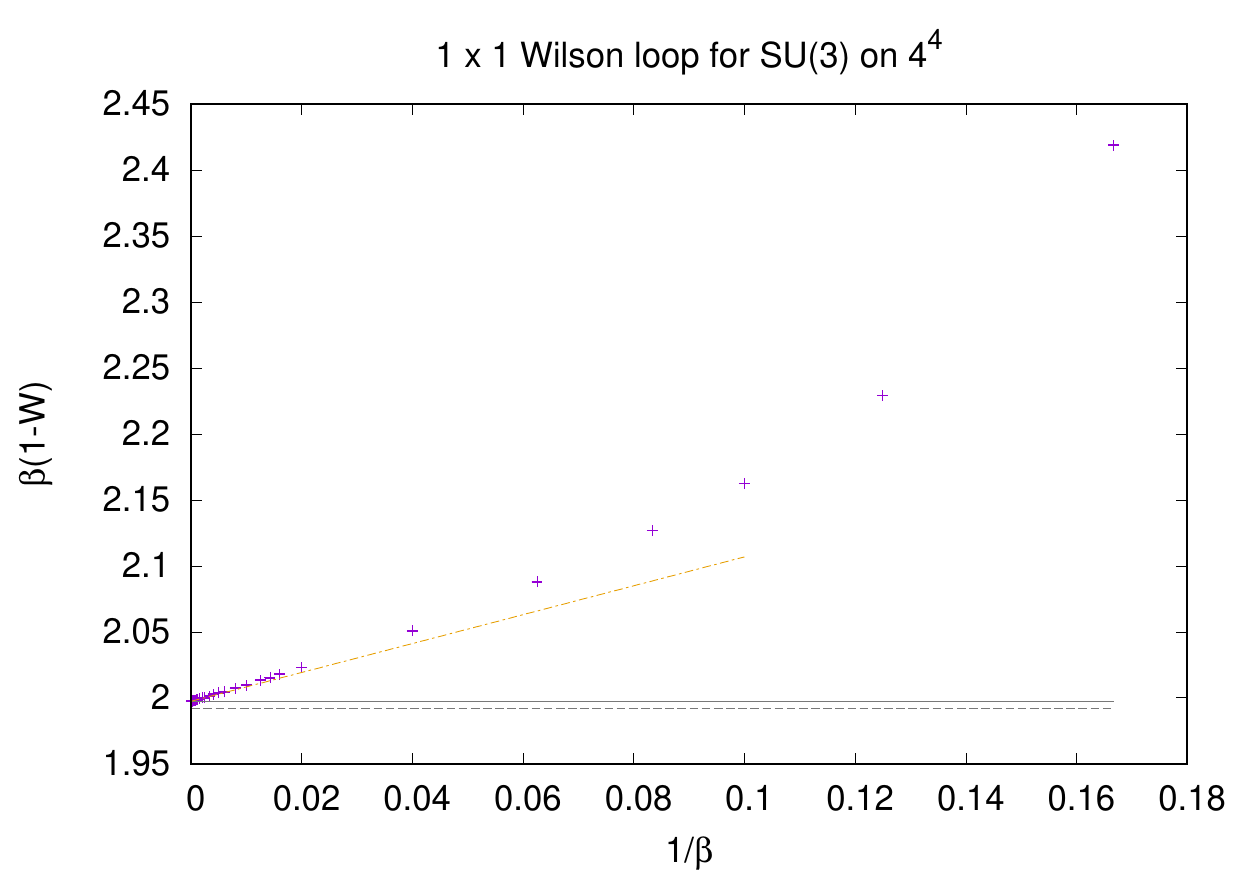}
 \includegraphics[scale=0.45]{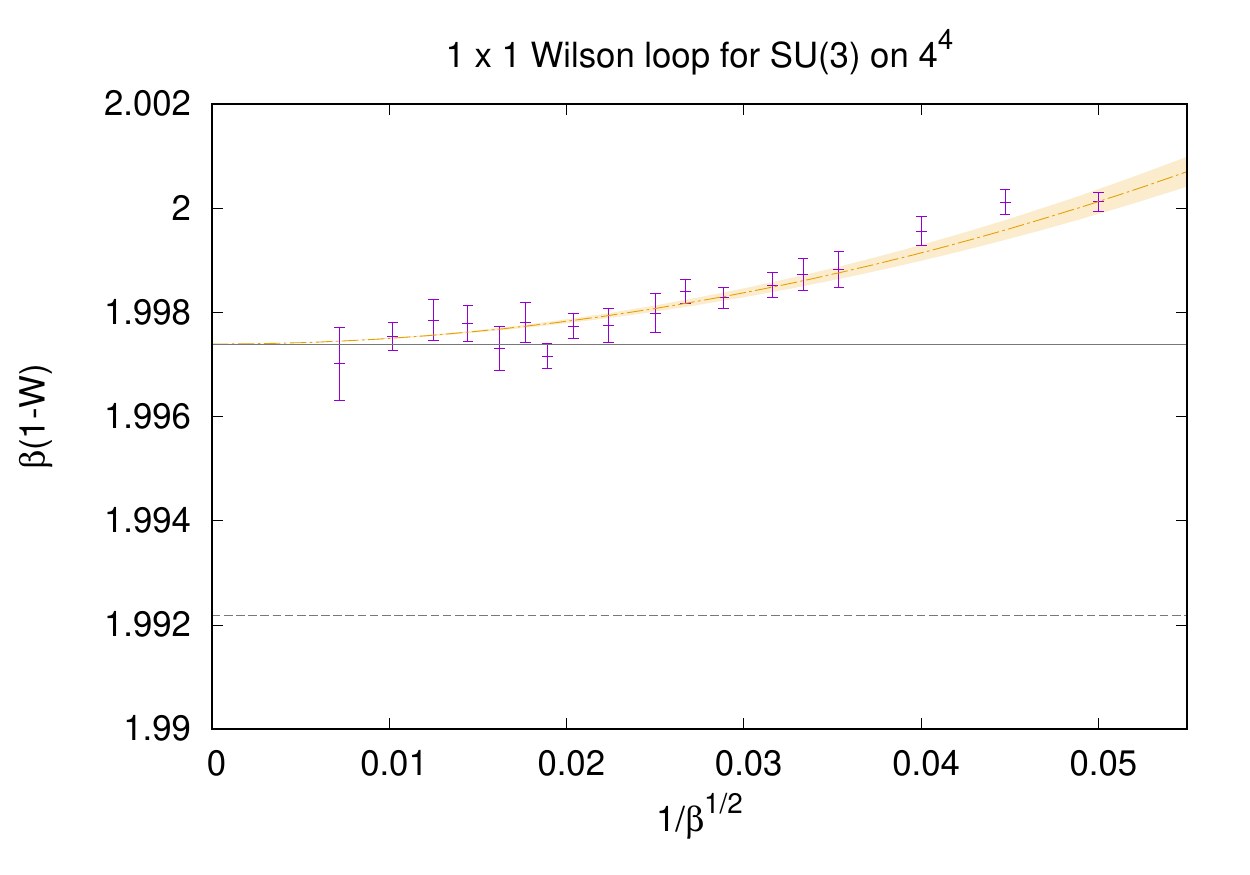}
 \includegraphics[scale=0.45]{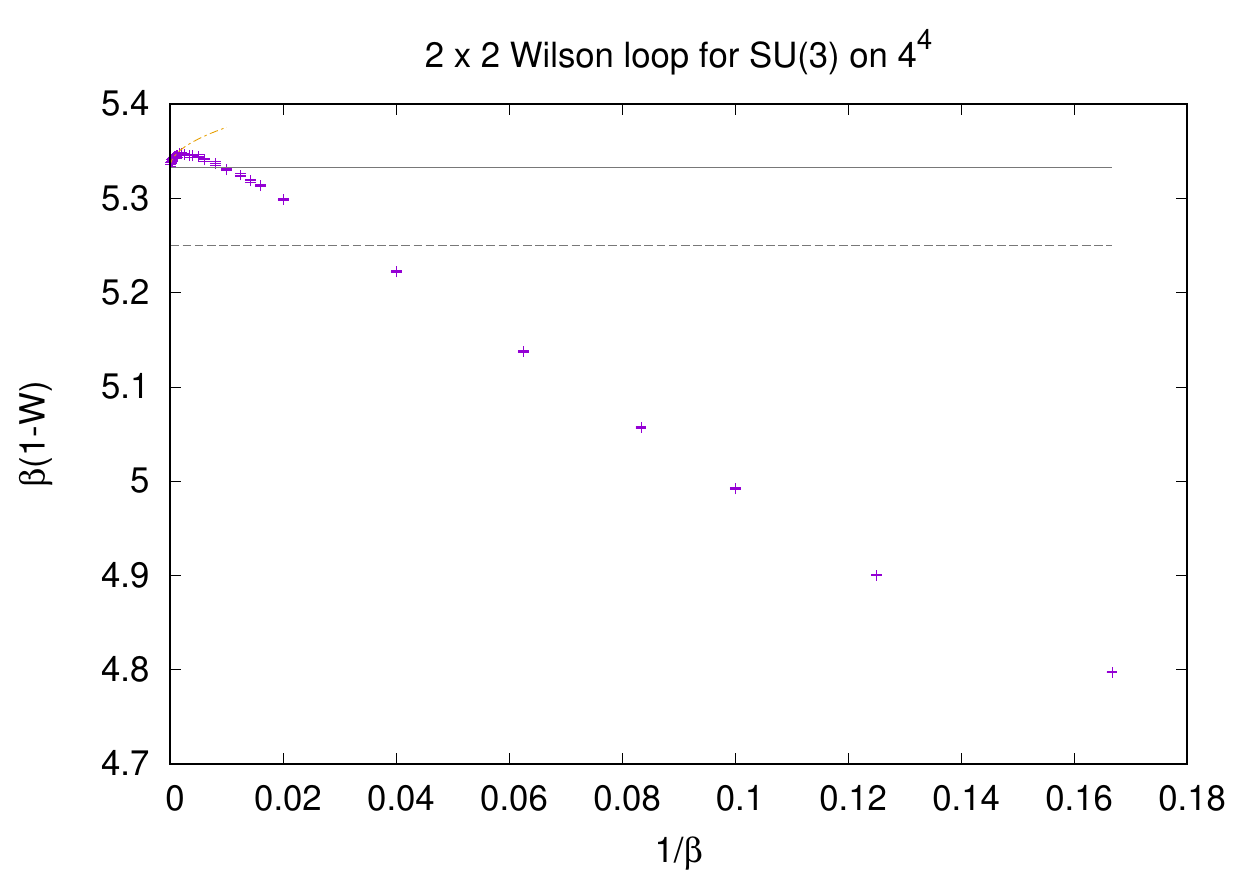}
 \includegraphics[scale=0.45]{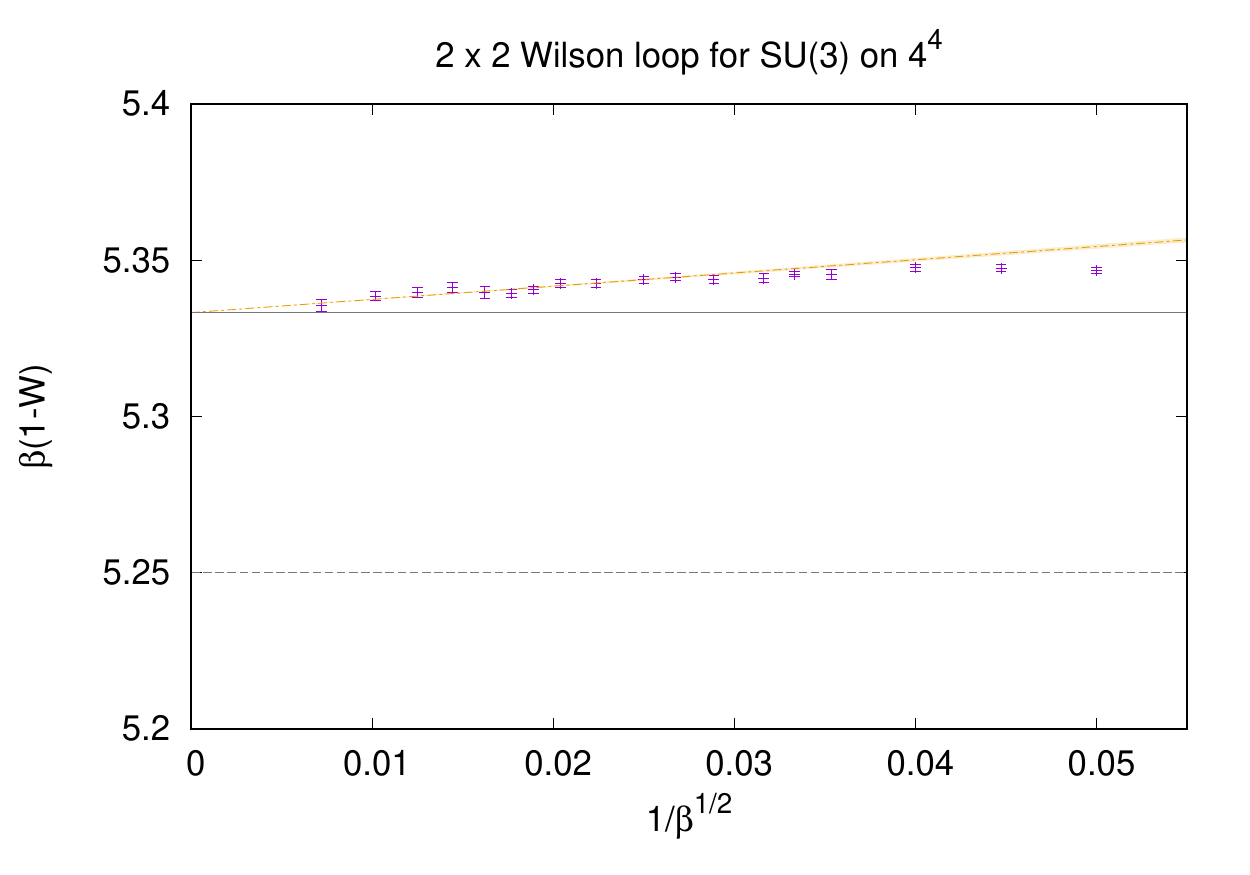}
 \includegraphics[scale=0.45]{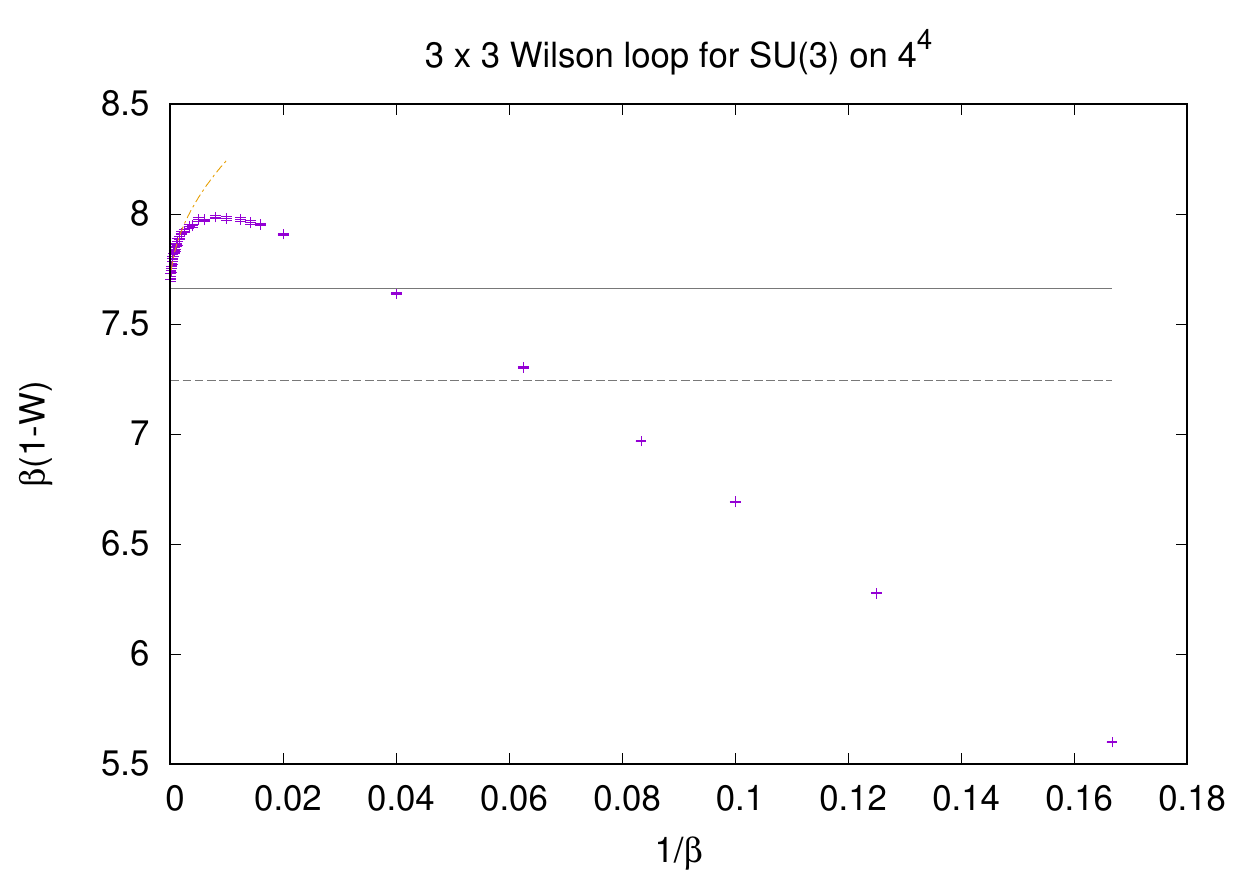}
 \includegraphics[scale=0.45]{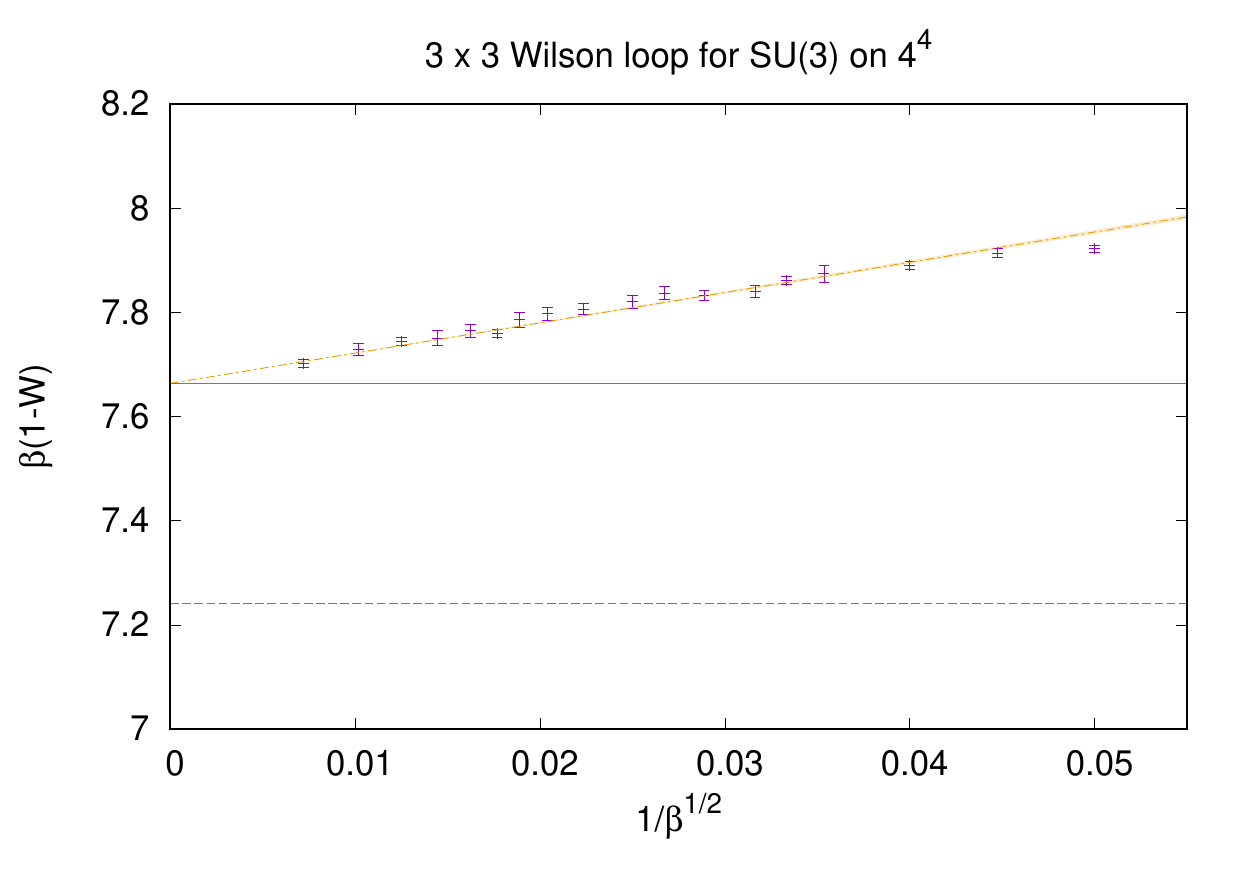}
 \includegraphics[scale=0.45]{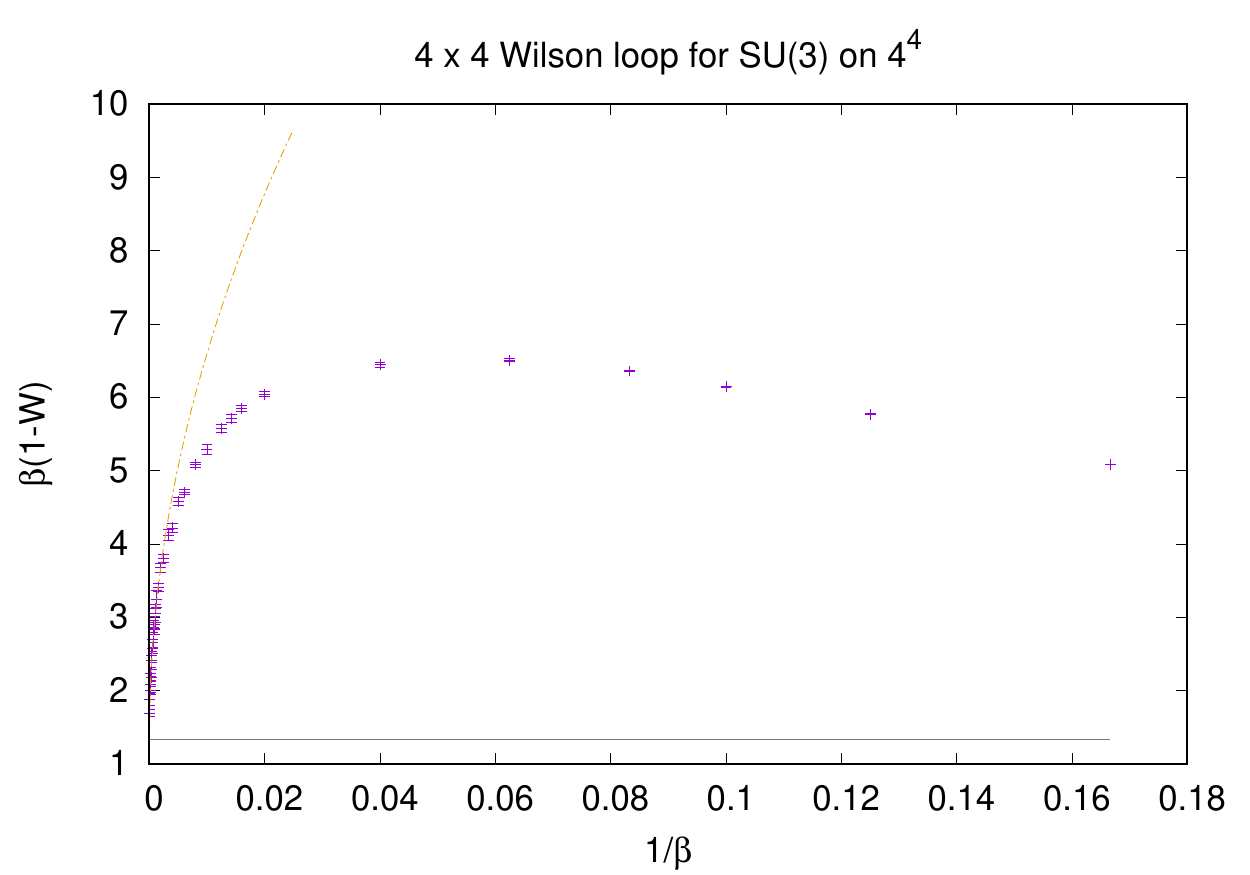}
 \includegraphics[scale=0.45]{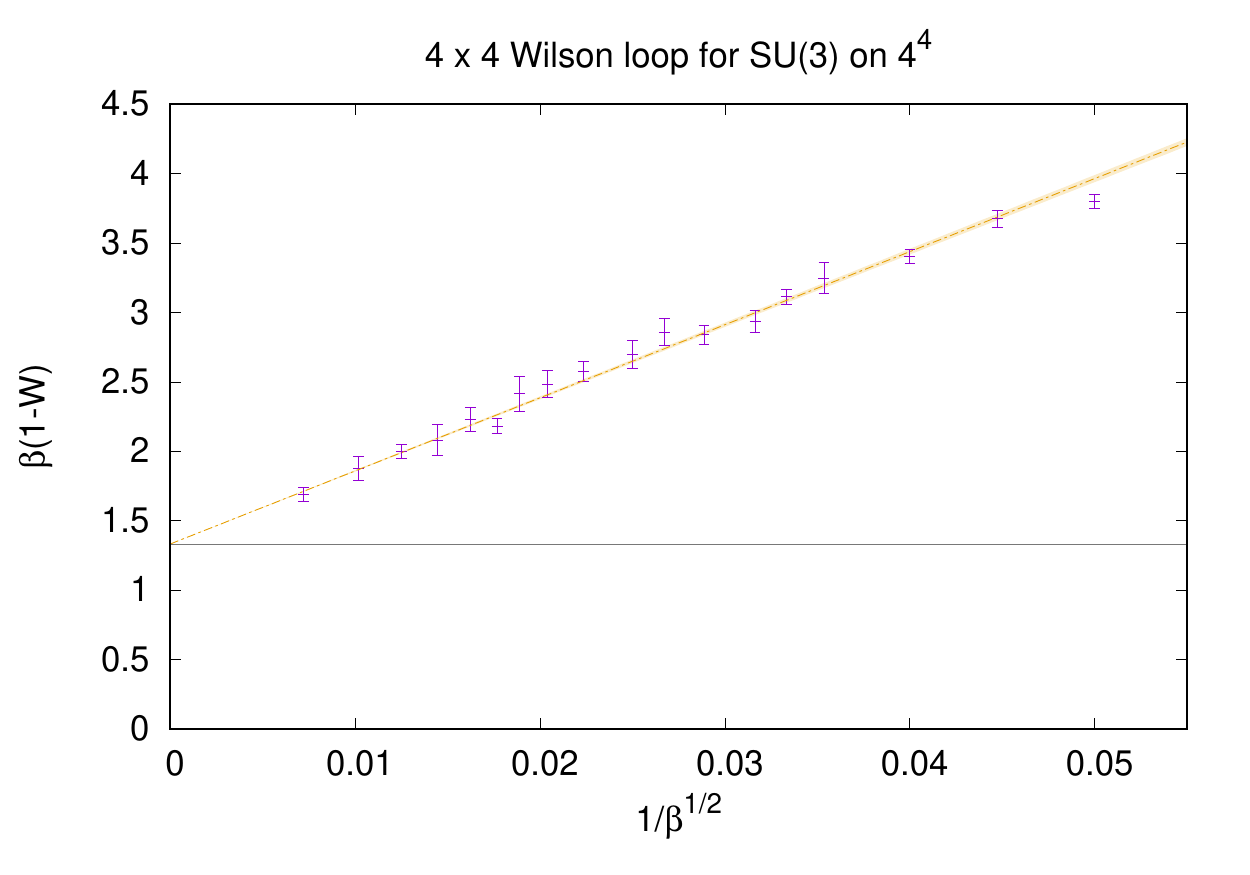}
 \caption{Monte Carlo results for $\beta \, \Big\{ 1-W(R_1,R_2) \Big\}$
   on a $4^4$ lattice
   are plotted against $\beta$ (Left) 
   and against $\beta^{-1/2}$ (Right) for $N=3$.
   The dashed and solid lines represent
   $c_1 ^{\rm (nonzero)}$ and
   $c_1^{\rm (tot)}$, respectively.
   Note that $c_1 ^{\rm (nonzero)}=c_1^{\rm (tot)}$
   for the $4\times 4$ maximal Wilson loop.
   The dash-dotted line represents a fit to the behavior
   $c_1^{\rm (tot)} + c \, \beta ^{-1/2}$
   except for the $1 \times 1$ Wilson loop,
   which is fitted by $c_1^{\rm (tot)} + c/\beta$.}
 \label{fig:b1-WL_Nc3_Ls4_Lt4}
\end{figure}

Let us first discuss the $N=3$ case.
In Fig.~\ref{fig:b1-WL_Nc3_Ls4_Lt4},
we plot
$\beta \, \Big\{ 1-W(R_1,R_2) \Big\}$
for square Wilson loops ($R_1=R_2$)
on a $4^4$ lattice with
$\beta= 6$\,--\,$19200$.
While the $1\times 1$ Wilson loop approaches $c_1^{\rm (tot)}$ monotonically,
the larger Wilson loops show prominent overshooting behaviors.
Moreover, the slope of the data points at the origin seems to be infinite
for the latter case, which suggests that the subleading terms become
non-analytic in $1/\beta$.
In fact, we can fit
$\beta \, \Big\{ 1-W(R_1,R_2) \Big\}$
to $c_1^{\rm (tot)} + c \, \beta ^{-1/2}$
as shown in Fig.~\ref{fig:b1-WL_Nc3_Ls4_Lt4} (Right)
although for the $1\times 1$ Wilson loop,
the coefficient $c$ is consistent with zero.
Therefore, 
the Wilson loops for the SU(3) case
are given by the expansion \eqref{WL-expansion},
where the subleading term
is
${\rm O}(\beta^{-\frac{3}{2}})$.
The coefficient of the non-analytic term $c \, \beta ^{-1/2}$
increases with the size of the Wilson loop,
which is responsible for the larger deviation 
from the leading perturbative prediction seen in 
Fig.~\ref{fig:b1-WL_Nc3_Ls4_Lt4} (Left).
This is reasonable since
the larger Wilson loops are more sensitive
to quantum fluctuations
in the infrared regime.
The origin of this non-analytic term can be understood from 
the viewpoint of the zero-mode fluctuations \cite{Coste:1985mn}
as we explain later.

\begin{figure}[H]
 \centering
 \includegraphics[scale=0.46]{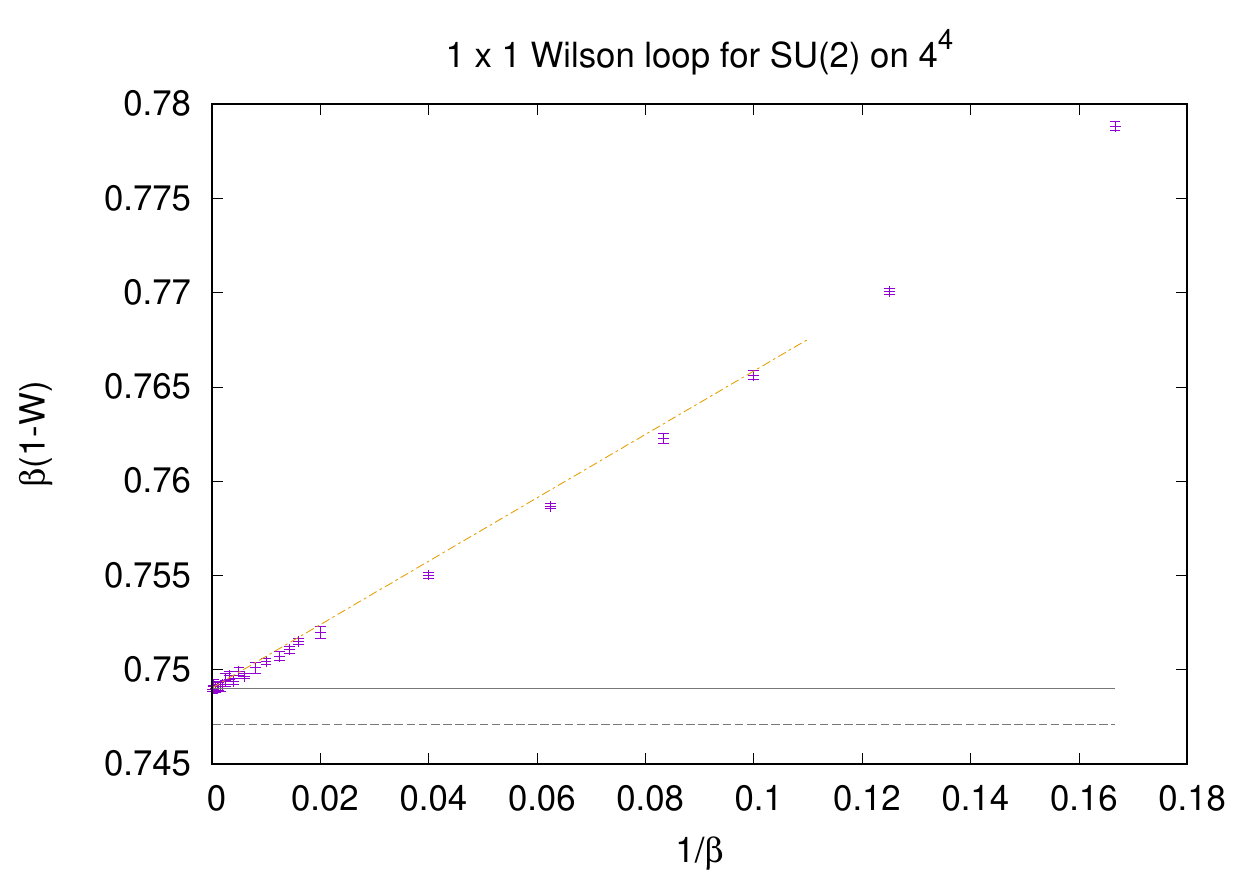}
 \includegraphics[scale=0.46]{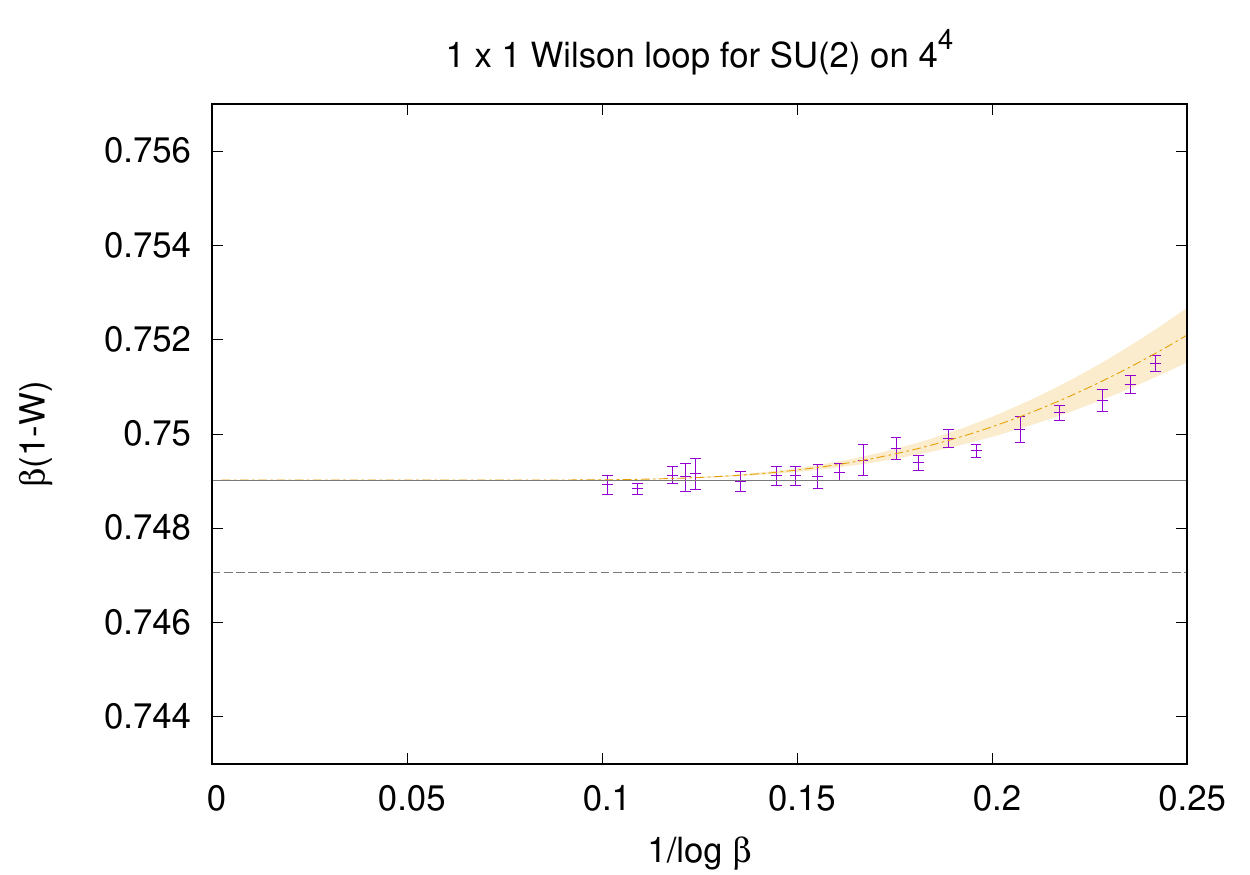}
 \includegraphics[scale=0.46]{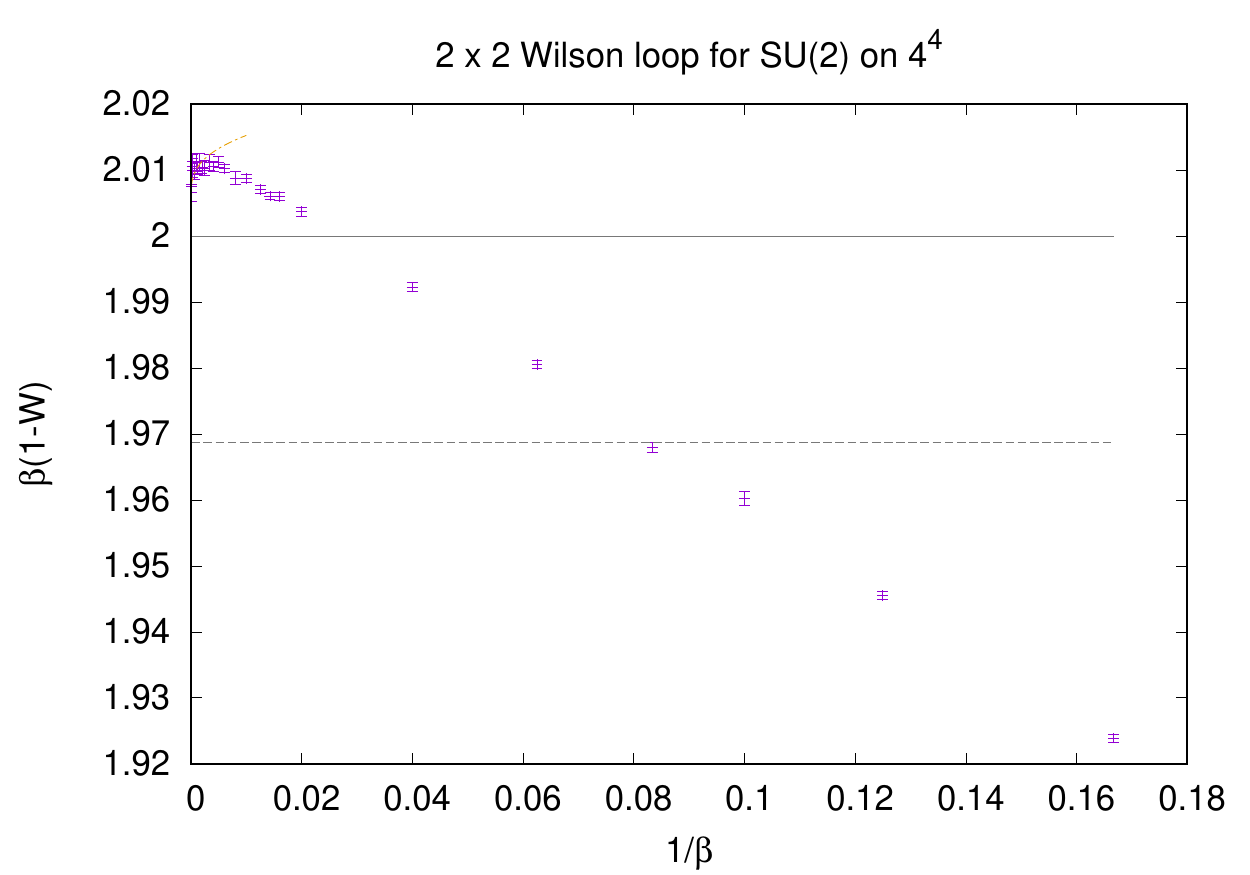}
 \includegraphics[scale=0.46]{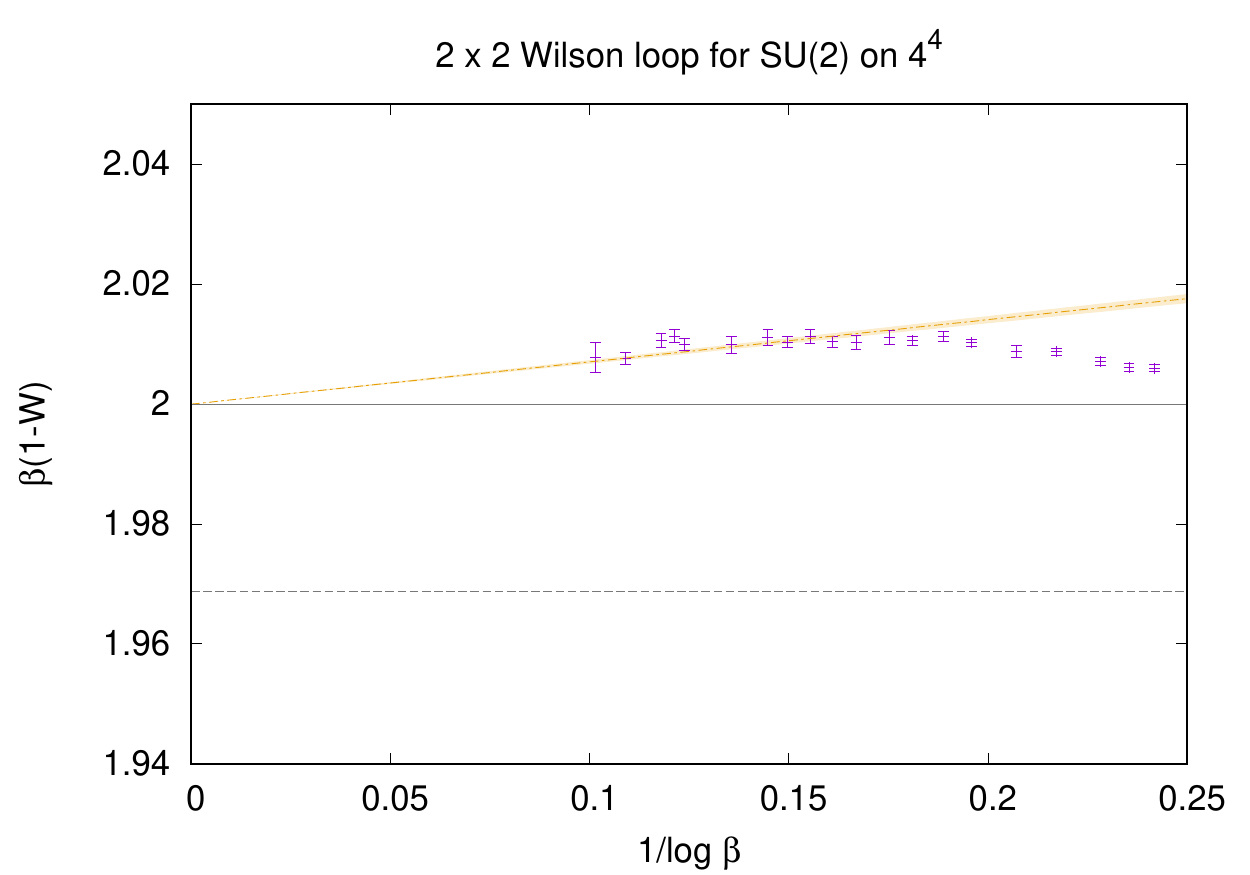}
 \includegraphics[scale=0.46]{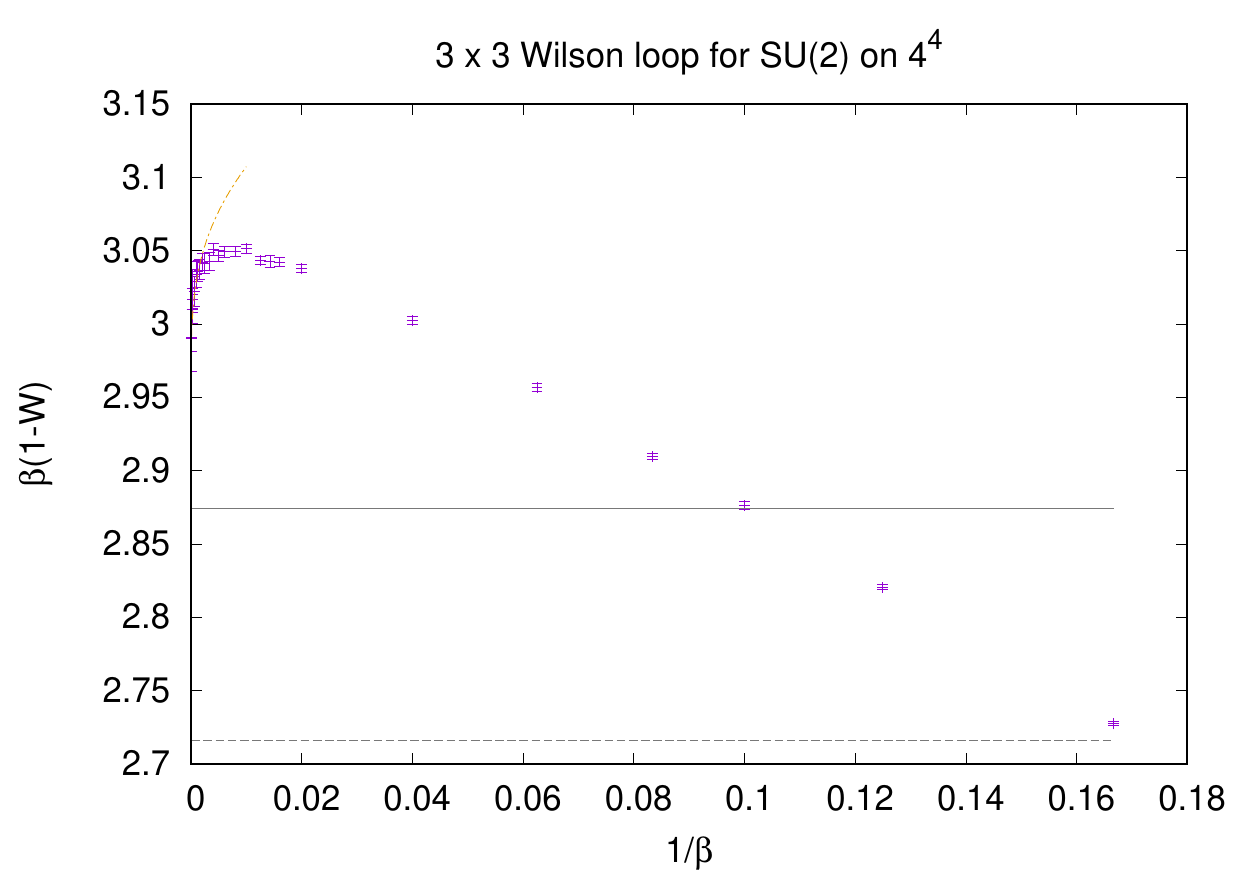}
 \includegraphics[scale=0.46]{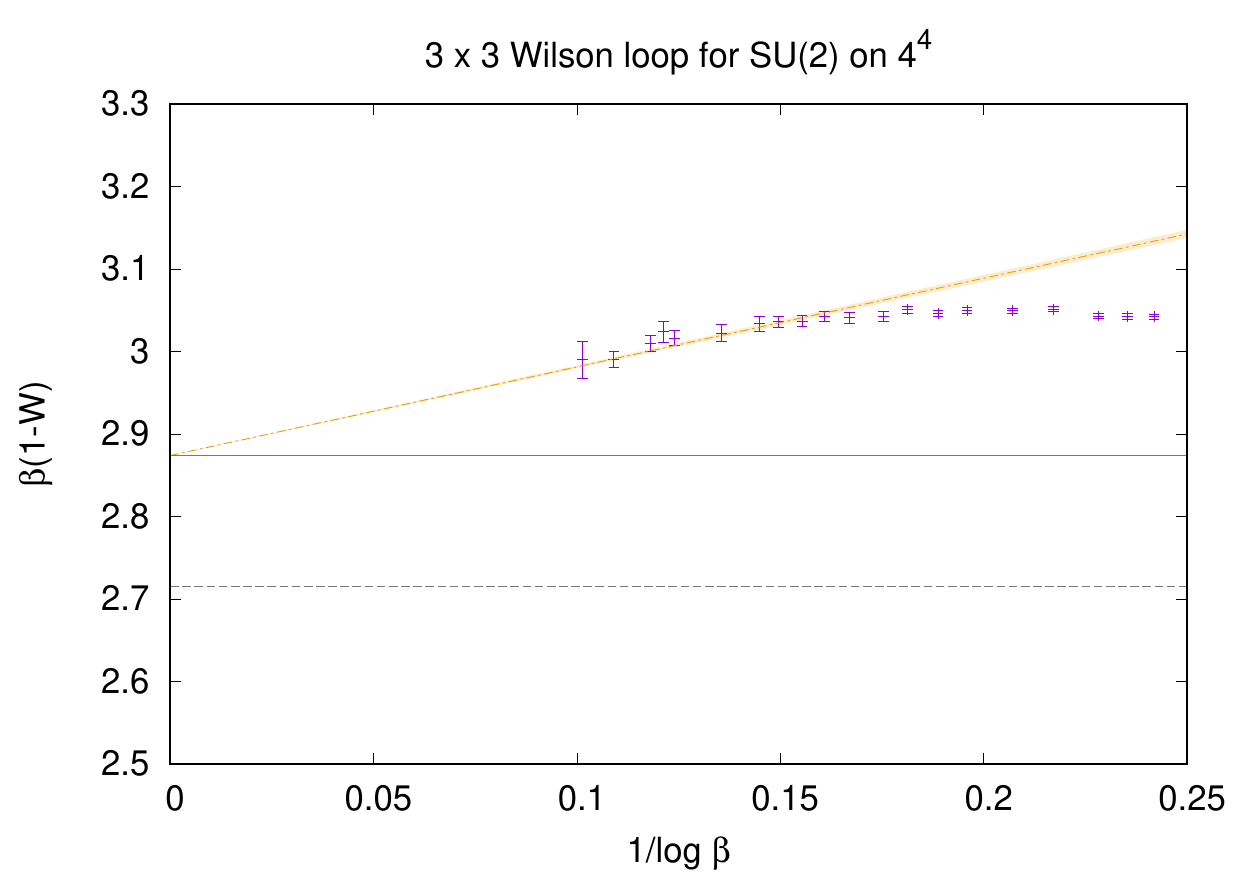}
 \includegraphics[scale=0.46]{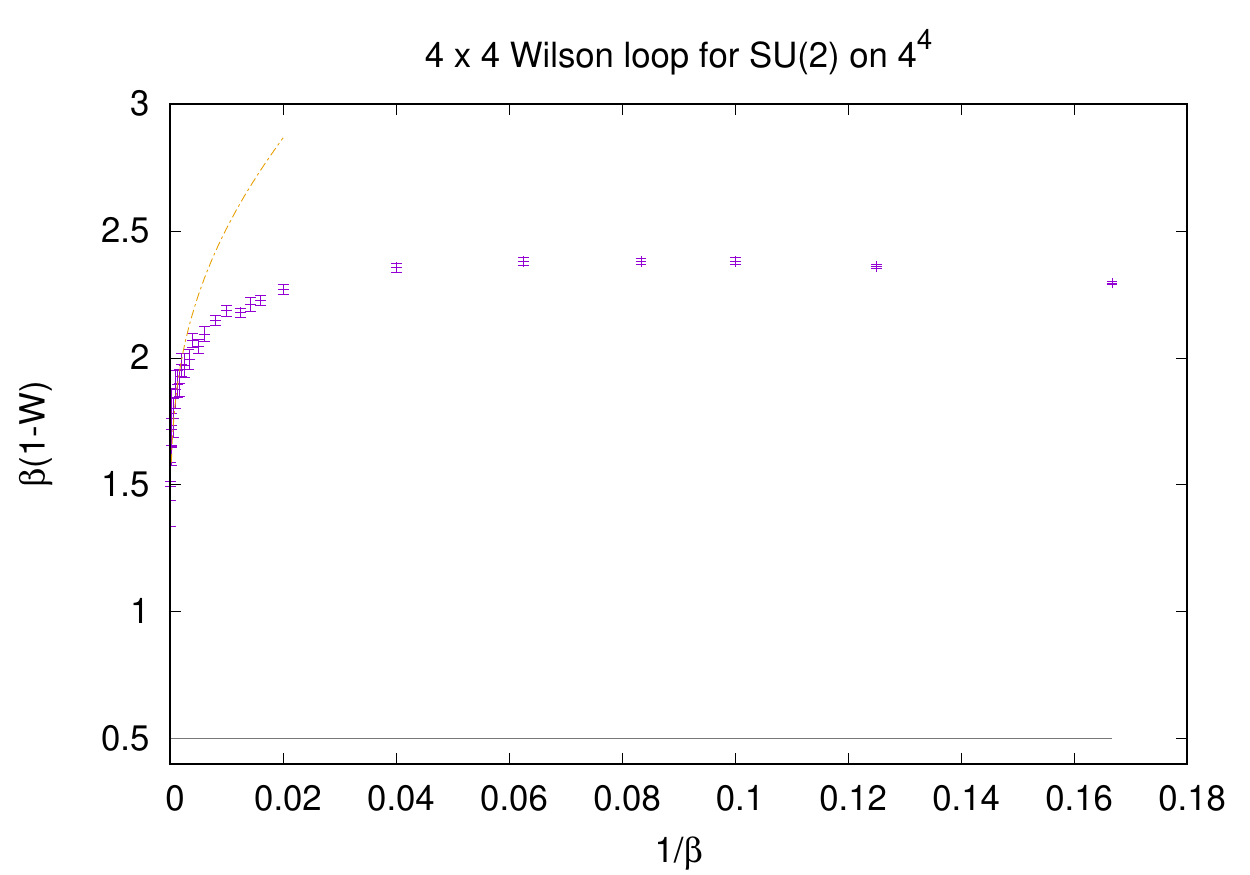}
 \includegraphics[scale=0.46]{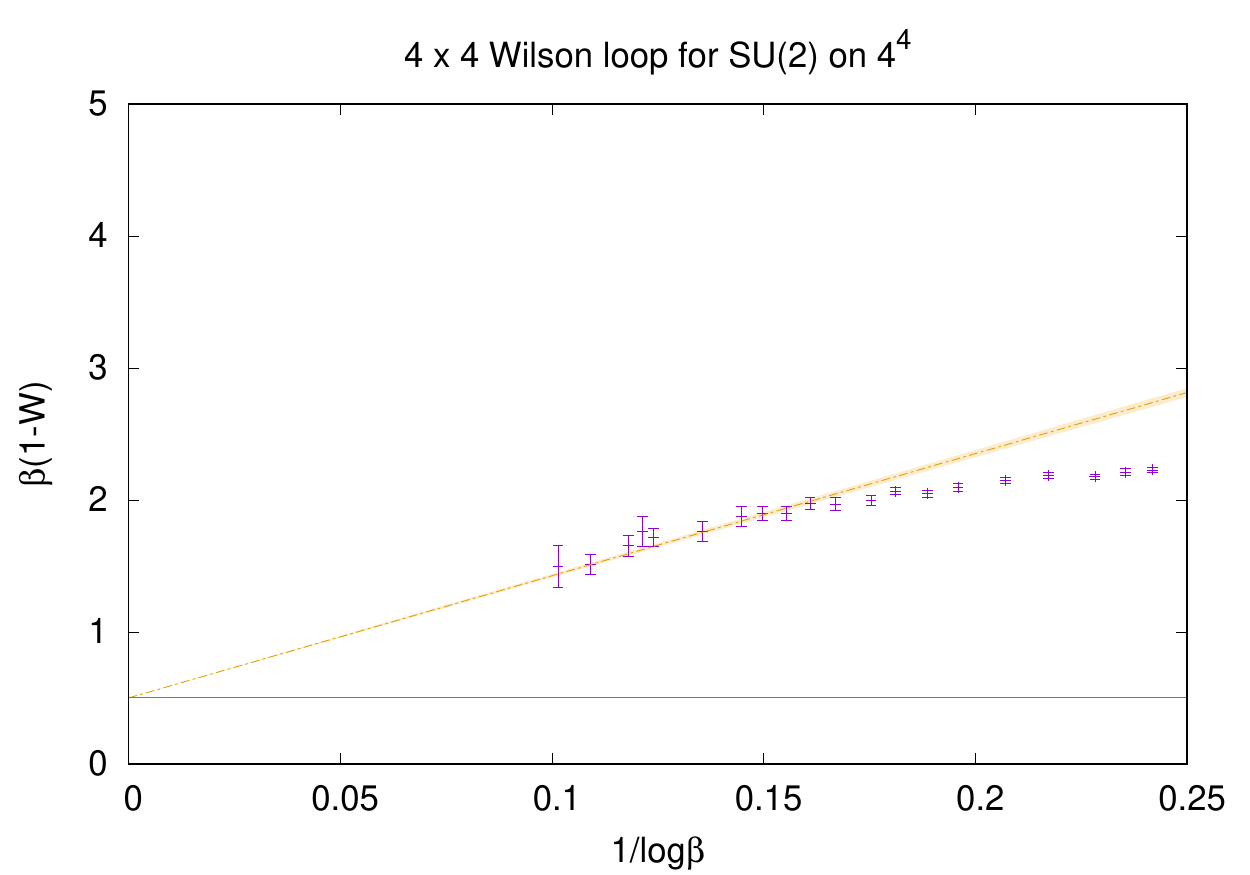}
 \caption{Monte Carlo results for $\beta \, \Big\{ 1-W(R_1,R_2) \Big\}$
   on a $4^4$ lattice
   are plotted against $\beta$ (Left) 
   and against $1/\log \beta$ (Right) for $N=2$.
   The dashed and solid lines represent
   $c_1 ^{\rm (nonzero)}$ and
   $c_1^{\rm (tot)}$, respectively.
   Note that $c_1 ^{\rm (nonzero)}=c_1^{\rm (tot)}$
   for the $4\times 4$ maximal Wilson loop.
   The dash-dotted line represents a fit to the behavior
$c_1^{\rm (tot)} + c / \log \beta$ except for the $1 \times 1$ Wilson loop,
   which is fitted by $c_1^{\rm (tot)} + c/\beta$.}
 \label{fig:b1-WL_Nc2_Ls4_Lt4}
\end{figure}

Next we discuss the $N=2$ case.
In Fig.~\ref{fig:b1-WL_Nc2_Ls4_Lt4},
we plot
$\beta \, \Big\{ 1-W(R_1,R_2) \Big\}$
for square Wilson loops ($R_1=R_2$)
on a $4^4$ lattice with
$\beta= 6$\,--\,$19200$.
From Fig.~\ref{fig:b1-WL_Nc2_Ls4_Lt4} (Left) alone,
it is not clear whether the data points
are approaching  $c_1^{\rm (tot)}$.
However, Fig.~\ref{fig:b1-WL_Nc2_Ls4_Lt4} (Right)
suggests that $\beta \, \Big\{ 1-W(R_1,R_2) \Big\}$
can be fitted to $c_1^{\rm (tot)}+c/\ln\beta$
although for the $1\times 1$ Wilson loop,
the coefficient $c$ is consistent with zero.
In particular, this result shows that the trivial vacuum
$U_{n,\mu}={\bf 1}$ dominates over the other toron configurations
at large $\beta$ even for the SU(2) case,
which is not obvious from Ref.~\cite{Coste:1985mn}.
We will provide clear understanding of this property
as well as the appearance of the logarithmic term $c/\ln\beta$
from the viewpoint of the zero-mode effective theory.

\section{Brief review of the zero-mode effective theory}
\label{sec:eff-theory}

In this section, we review
the zero-mode effective theory following Ref.~\cite{Coste:1985mn}.
We consider the classical solutions, which are given by toron configurations,
and discuss the perturbative expansion around them including the zero modes.
In particular, we find in the SU(3) case
that the trivial vacuum dominates at large $\beta$ and
that the leading term in the Wilson loop
can be obtained by using the zero-mode effective theory.



\subsection{The effective action for the zero modes}
The minima of the plaquette action are degenerate
under periodic boundary conditions,
and they are given,
up to a gauge transformation,
by torons $U_{n,\mu}^{(0)}=e^{iB_\mu}$, where
$B_\mu$ is a constant real diagonal matrix.
Let us
expand the link variables around them
as
\begin{align}
  U_{n,\mu}=e^{iQ_{n,\mu}} \, e^{iB_\mu} \ ,
  \label{U-expand}
\end{align}
where $Q_{n,\mu}$ represents the fluctuations
given by traceless Hermitian matrices satisfying
\begin{align}
  \frac{1}{V}\sum_n (Q_{n,\mu})_{ii} = 0 \quad \quad  \mbox{for all $i$} \ .
 \label{zero-mode_cond}
\end{align}
Plugging \eqref{U-expand} into \eqref{plaq-action}, we obtain
\begin{align}
 S_{\rm g}
 &=-\frac{\beta}{2N}\sum_n\sum_{\mu,\nu\ne\mu}
 \tr \Big( e^{iQ_{n,\mu}}e^{i(1+D_\mu)Q_{n,\nu}}e^{-i(1+D_\nu)Q_{n,\mu}}e^{-iQ_{n,\nu}}-1 \Big) \ ,
\label{action-QB}
\end{align}
where $D_\mu Q_{n,\nu}=e^{i\adj(B_\mu)}Q_{n+\hat\mu,\nu}-Q_{n,\nu}$
and $\adj(B_\mu)=[B_\mu, \ \cdot \ ]$.

Let us then decompose $Q_{n,\mu}$ as
\begin{align}
  Q_{n,\mu}=A_\mu+ \tilde{A}_{n,\mu} \ ,
\end{align}
where $A_\mu$ represents the zero modes,
which satisfy $(A_\mu)_{ii}=0$ for all $i$
due to \eqref{zero-mode_cond}.
The plaquette that appears in \eqref{action-QB}
can be expressed as
\begin{align}
 &e^{iQ_{n,\mu}}e^{i(1+D_\mu)Q_{n,\nu}}e^{-i(1+D_\nu)Q_{n,\mu}}e^{-iQ_{n,\nu}}
 \nonumber \\
 &=\exp\bigg[ 
i \bigg\{
   \Big(D_\mu \tilde{A}_{n,\nu} - D_\nu \tilde{A}_{n,\mu} \Big)
   +  
\Big(D_\mu A_{\nu}
 -D_\nu A_{\mu}\Big)  +i \Big[A_{\mu}, A_{\nu} \Big]
 \nonumber \\
 &\hspace{70pt}
 +i \left[A_{\mu},D_\mu A_{\nu}-\frac{1}{2}D_\nu A_{\mu} \right]
 +i \left[D_\nu A_{\mu}-\frac{1}{2}D_\mu A_{\nu},A_{\nu}\right]
 \nonumber \\
 &\hspace{70pt}
 +\frac{i}{2} \Big[D_\nu A_{\mu},D_\mu A_{\nu}\Big]
 +\cdots
 \bigg\} \bigg] \ , 
 \label{plaquette-expansion}
\end{align}
where $D_\mu A_\nu=(e^{i\adj(B_\mu)}-1) \, A_\nu$.
Thus, the action 
can be rewritten as
\begin{align}
 S_{\rm g}
 &=\frac{1}{2N}\sum_{\mu,\nu\ne\mu}\tr\bigg[
 \frac{\beta}{2}\sum_n \Big(D_\mu \tilde{A}_{n,\nu} - D_\nu \tilde{A}_{n,\mu} \Big)^2
 - \beta V \Big( U_{\mu}U_{\nu}U_{\mu}^\dagger U_{\nu}^\dagger - 1 \Big)
 +{\rm O} \Big( (\beta V)^{-\frac{1}{4}} \Big)
 \bigg] \ ,
 \label{Sg-expansion}
\end{align}
where we have defined
\begin{align}
U_\mu=e^{iA_\mu} e^{iB_\mu} \ .
\label{def-U-zero-mode}
\end{align}
From this, one finds that the elements of $A_\mu$
are generically of the order of 
$(\beta V)^{-\frac{1}{2}}$.
Care should be taken when there exist $i$ and $j\neq i$
such that $(B_{\nu})_{ii}- (B_{\nu})_{jj}=0$ for any $\nu$.
In that case, one has
$\Big (\adj(B_\nu)A_\mu \Big)_{ij}=0$ for any $\nu$.
Therefore, one finds that the $ij$ element of $A_\mu$ is of the
order of $( \beta V) ^{-\frac{1}{4}}$.
Based on this power counting\footnote{Note
that this power counting is not valid for the 4D SU(2) case,
where the fluctuations in $A_\mu$ are larger than O($(\beta V)^{-1/4}$).},
one finds that
the terms abbreviated as $\cdots$ in \eqref{plaquette-expansion}
are of the order of $(\beta V)^{-\frac{1}{4}}$ 
at most.
One can rewrite the Haar measure $dU_{n,\mu}$
in terms of $dB_\mu$, $dA_\mu$ and $d\tilde{A}_{n,\mu}$ as
\begin{align}
dU_{n,\mu}
 &=dB_\mu \, dA_\mu d\tilde{A}_{n,\mu}
 \Big(1+{\rm O}(V^{\frac{1}{2}}\beta^{-\frac{1}{2}})\Big) \ .
 \label{Haar-measure}
\end{align}
\mycomment{The ${\rm O}(V^{\frac{1}{2}}\beta^{-\frac{1}{2}})$ term
in \eqref{Haar-measure}
becomes ${\rm O}(\beta^{-1})$ if $B_\mu$ is not degenerate.
Therefore, the measure term does not contribute in the action to leading order.}

Let us introduce the gauge fixing term for the non-zero modes
\begin{align}
 S_{\rm g.f.}
 &=\frac{\beta}{2N} 
 \sum_n\tr\left[\left( \sum_\mu D_\mu \tilde{A}_{n,\mu} \right)^2 \right]
 =\frac{\beta}{2N}\sum_{k\ne 0}\tr\left|
 \sum_{\mu}
 (e^{i(k_\mu+\adj(B_\mu))}-1)
 \tilde{A}_\mu(k)
 \right|^2 \  ,
 \label{g-f-term}
\end{align}
where $\tilde{A}_\mu(k)$ represents the Fourier modes of $\tilde{A}_{n,\mu}$.
At the leading order in $1/\beta$,
the partition function becomes
\begin{align}
 Z = \int  dB_\mu \, dA_\mu d\tilde{A}_{n,\mu} \, dc \, d\bar{c}
 \, e^{-S_\text{zero}-S_{\rm nonzero}} 
 \ ,
\end{align}
where $c$ and $\bar{c}$ are the ghosts
associated with the gauge fixing \eqref{g-f-term}
and the actions for the zero modes and the nonzero modes
are given, respectively, as
\begin{align}
S_\text{zero}
  &= -\frac{\beta V}{2N}\sum_{\mu,\nu\neq\mu}
\tr \Big( U_\mu U_\nu U_\mu^\dagger U_\nu^\dagger - 1 \Big) \ ,
\label{action-zero-mode}
 \\
 S_{\rm nonzero} &= \frac{\beta}{2N}\sum_{k\ne 0} \sum_{\mu,\lambda}
 4\tr  \left\{ \tilde{A}_\mu(-k)
 \sin^2\frac{k_\lambda+\adj(B_\lambda)}{2}
 \tilde{A}_\mu(k) \right.
 \nonumber
 \\
 & \quad\quad\quad\quad\quad\quad\quad\quad\quad
 \left.  +\bar{c}(-k)
 \sin^2\frac{k_\lambda+\adj(B_\lambda)}{2}
 c(k) \right\}
 \ .
 \label{action-nonzero-mode}
\end{align}

The expectation value of an observable $\mathcal{O}$ can be obtained as
\begin{align}
 \left\langle \mathcal{O} \right\rangle
 =\frac{1}{Z}\int dB_\mu \, dA_\mu d\tilde{A}_{n,\mu} \,  dc \, d\bar{c}
 \, \mathcal{O} \, 
 e^{-S_\text{zero}-S_{\rm nonzero}} 
 \ .
 \label{ev-leading}
\end{align}

\subsection{The integration over the torons}
\label{sec:int-torons}

In this section, we discuss the integration over the torons $B_\mu$,
which represent the minimum action configurations.

As stated below \eqref{Sg-expansion},
the order of magnitude of
the zero modes $A_\mu$
depends on
the toron configuration $B_\mu$.
Here we parametrize the toron configuration as
\begin{align}
 B_\mu^{(0)}
 =
 \begin{pmatrix}
b_\mu^{(1)}\mathbf{1}_{V_1}  &  &  &  \\
               & b_\mu^{(2)}\mathbf{1}_{V_2}  &  &   \\
  &  &  \ddots  &  \\
 & & & b_\mu^{(M)}\mathbf{1}_{V_M}\\
\end{pmatrix} \ ,
 \label{barB-def}
\end{align}
where $(V_1,\cdots,V_{M})$ is a partition of $N$,
and consider the fluctuations
$B_\mu=B_\mu^{(0)} +  \tilde{B}_\mu$
around \eqref{barB-def}.
%
The plaquette appearing in the action \eqref{action-zero-mode}
for the zero modes can then be expanded as
\begin{align}
  &\sum_{\mu \ne \nu}
  \tr \Big(
  e^{iQ_{n,\mu}}e^{i(1+D_\mu)Q_{n,\nu}}e^{-i(1+D_\nu)Q_{n,\mu}}e^{-iQ_{n,\nu}}-1 \Big)
 \nonumber \\
 &=-\frac{1}{2}\tr\bigg[ 
 \bigg\{
 i (
 [B_\mu^{(0)}, A_{\nu}]
 -[B_\nu^{(0)}, A_{\mu}]
 )
 +i[\tilde{B}_\mu, A_{\nu}]
 -i[\tilde{B}_\nu, A_{\mu}]
 +i[A_{\mu}, A_{\nu}]
 \nonumber \\
 &\hspace{70pt}
 -\left[A_{\mu},[B_\mu^{(0)}, A_{\nu}]
   -\frac{1}{2}[B_\nu^{(0)}, A_{\mu}]\right]
 -\left[[B_\nu^{(0)}, A_{\mu}]
   -\frac{1}{2}[B_\mu^{(0)}, A_{\nu}],A_{\nu}\right]
 \nonumber \\
 &\hspace{70pt}
 -\frac{i}{2}\Big[ [B_\nu^{(0)}, A_{\mu}],[B_\mu^{(0)}, A_{\nu}] \Big]
 \bigg\}^2 +\cdots \bigg] \  .
\end{align}
Thus, the elements of $A_\mu$
that appear in the action quadratically are
those outside the $V_r\times V_r$ blocks.
The number of such modes is
\begin{align}
 E_2(V)=
 (D-1)\left(
 N^2 - \sum_{r=1}^{M} V_r^2 \right) \  ,
\end{align}
where $-1$ in the factor $(D-1)$ is due to
the gauge fixing of the zero modes \cite{Coste:1985mn}.
On the other hand,
the elements of $A_\mu$
that appear quartically
are those inside the $V_r\times V_r$ blocks.
The number of such modes is
\begin{align}
 E_4(V)=
 D\sum_{r=1}^{M}(V_r^2-1) \  ,
\end{align}
after subtracting the number of modes corresponding to $B_\mu^{(0)}$.
Each of the quadratic and quartic modes contribute to the
partition function a factor of $O(\beta^{-1/2})$ and $O(\beta^{-1/4})$,
respectively.
Hence, the fluctuations around the toron labeled by $(V_1,\cdots,V_M)$ 
contribute to the partition function as
\begin{align}
Z(V) \sim   \beta^{-\{\frac{1}{2}E_2(V)+\frac{1}{4}E_4(V)\}} \  ,
\end{align}
which implies that
the toron configurations with $V$
that minimizes 
$\frac{1}{2}E_2(V)+\frac{1}{4}E_4(V)$
give the dominant contribution.
From this argument, one finds a critical dimension
\begin{align}
D_c =\frac{2N}{N-1} 
\label{critical-D}
\end{align}
such that
for $D\le D_c$,
the dominant configuration corresponds to
$M=N$ with $V_1=V_2=\cdots =V_{N}=1$,
while for $D\ge D_c$,
it corresponds to $M=1$ with $V_1=N$.

In $D=4$ dimensions, in particular, 
$N=3$ corresponds to the latter case,
which means that the trivial vacuum $B_\mu=0$
dominates in the SU(3) case.
On the other hand, $N=2$ is marginal
and one cannot determine the dominant toron configurations
from this argument alone in the SU(2) case.\footnote{According to 
this argument, 
nontrivial toron configurations dominate over the trivial vacuum
at large $\beta$ in the case of 3D SU(2) gauge theory and 
2D SU($N$) gauge theories with arbitrary $N$.\label{footnote:3d-SU2}}

\subsection{Perturbative expansion of the Wilson loops}
The discussion above implies that one can obtain the perturbative
expansion of the Wilson loops for $N=3$.
Since the dominant toron is the trivial vacuum,
$B_\mu$ fluctuates around 0;
namely
$B_\mu=\tilde{B}_\mu$.
Expanding the zero-mode effective theory \eqref{action-zero-mode}
with respect to $A_\mu$ and $\tilde{B}_\mu$ using \eqref{def-U-zero-mode},
we obtain, at the leading order,
\begin{align}
Z_0 
&= \int dX_\mu \, e^{ - S_0} \ , \nonumber \\
S_0 &= - \frac{\beta V}{4N} \sum_{\mu \ne \nu} \tr [X_\mu , X_\nu]^2 \ ,
\label{reducedYM-action}
\end{align}
where $X_\mu = A_\mu + \tilde{B}_\mu$.
Thus, 
the zero-mode contribution
to the Wilson loop in \eqref{WL-expansion}
can be obtained by
\begin{align}
 -\frac{c_1^{\rm (zero)}}{\beta}
 &= \frac{1}{2N} (R_1 R_2)^2
 \Big\langle \tr \Big(
 [ X_{1} , X_{2} ]^2
 \Big) \Big\rangle_0  \ ,
\end{align}
where the expectation value
$\langle \ \cdot \ \rangle_0$
is taken with respect to \eqref{reducedYM-action}.
Using the Schwinger-Dyson equation
\begin{align}
  -\frac{\beta V}{N}
  \sum_{\mu \ne \nu}
  \Big\langle\tr 
 [X_\mu, X_\nu]^2  \Big\rangle
 = D \, (N^2-1) \  ,
 \label{SDeq_for_zero-mode}
\end{align}
derived from \eqref{reducedYM-action},
one obtains \eqref{c1-0}.

\mycomment{
One can, however, put an upper bound and a lower bound
on the partition function, which implies that the
partition function diverges as
$\log \beta$ at large $\beta$ \cite{Coste:1985mn}.
}

\mycomment{
The next to the order of $\beta^{-1}$ is $\beta^{-\frac{3}{2}}$ in general
because in the expansion enter an even number of the zero modes,
which are of ${\rm O}(\beta^{-\frac{n}{2}})$ with integer $n$
but there are no terms of ${\rm O}(\beta^{-\frac{2n+1}{4}})$.
This is how the Wilson loops are perturbatively expanded in terms of $\beta^{-\frac{1}{2}}$,
in the case of four-dimensional SU(3) lattice QCD.
}

\section{Understanding based on the reduced model}
\label{sec:reduced-YM}

In this section, we show
how the difference between SU(2) and SU(3) in 4D
can be understood
in terms of the zero-mode effective theory.
For that, we have performed Hybrid Monte Carlo simulations of
the zero-mode effective theory 
\eqref{reducedYM-action}
for $N=2$ and $N=3$ with $D=4$,
where the coefficient $\beta V$ in the action
is set to unity since it can be absorbed by 
rescaling $X_\mu$.

\begin{figure}[htbp]
 \centering
 \includegraphics[scale=0.62]{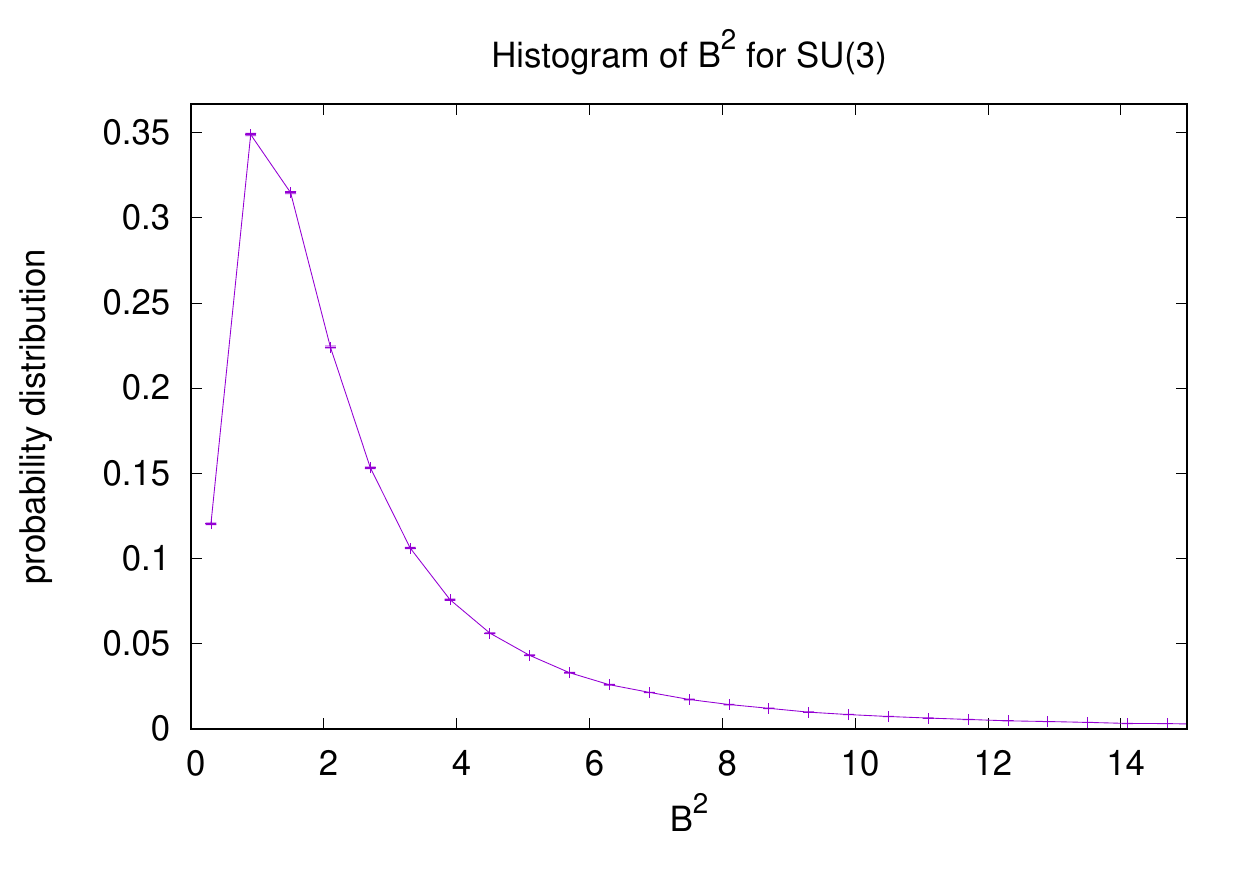}
 \includegraphics[scale=0.62]{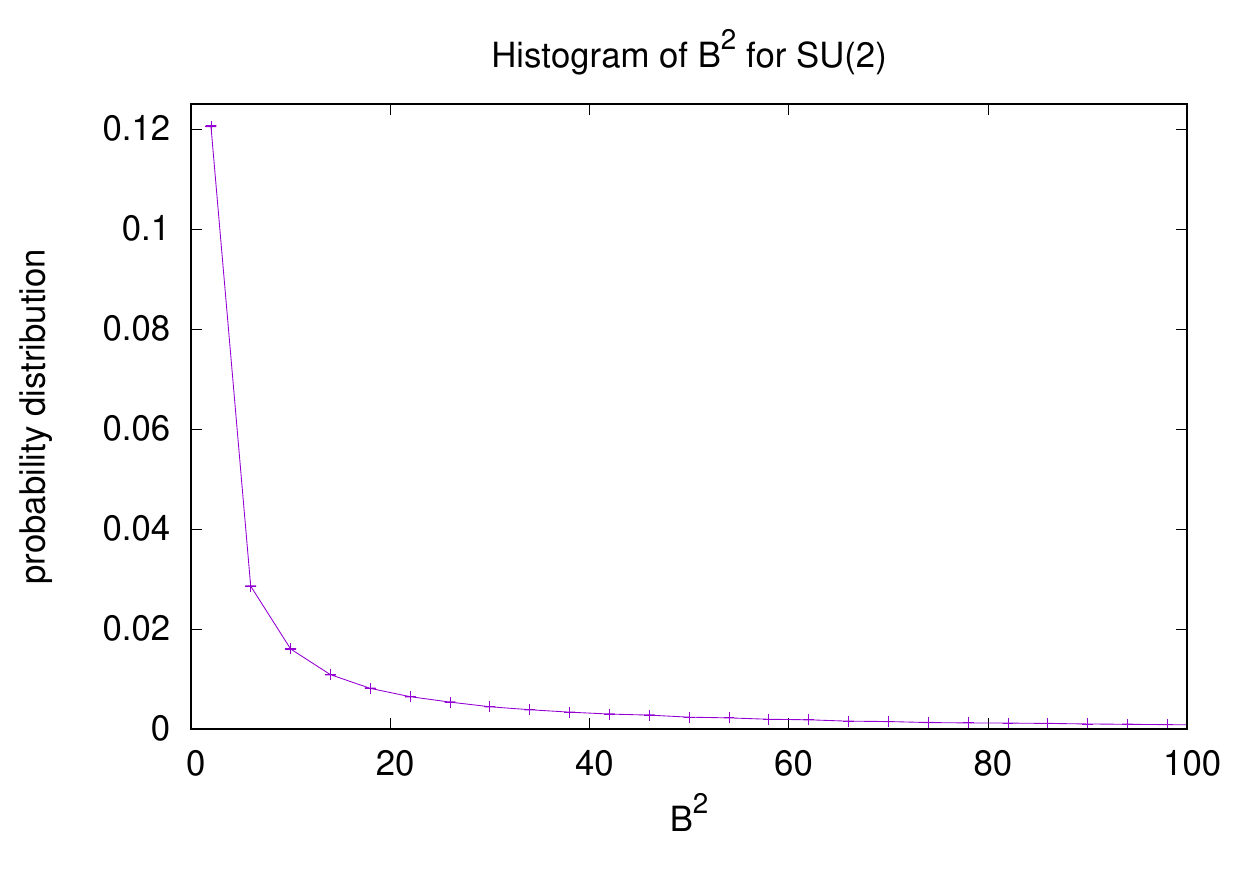}
 \caption{Histograms of $\frac{1}{N}\tr (B_\mu)^2$ 
obtained by simulating the zero-mode effective theory 
for $N=3$ (Left) and $N=2$ (Right).
 }
 \label{fig:Bsq_for_zero-mode}
\end{figure}

In Fig.~\ref{fig:Bsq_for_zero-mode}
we plot
the histogram 
of $\tr (\tilde{B}_\mu)^2$ 
for $N=2$ and $N=3$.
We observe a peak near the origin for both cases,
which suggests that the trivial vacuum dominates at large $\beta$
not only for SU(3) but also for SU(2).
However, the difference is that the 
histogram
for the SU(2) case
has a longer tail than that for the SU(3) 
case (Notice the scale of the horizontal axis.).

\begin{figure}[htbp]
 \centering
 \includegraphics[scale=0.62]{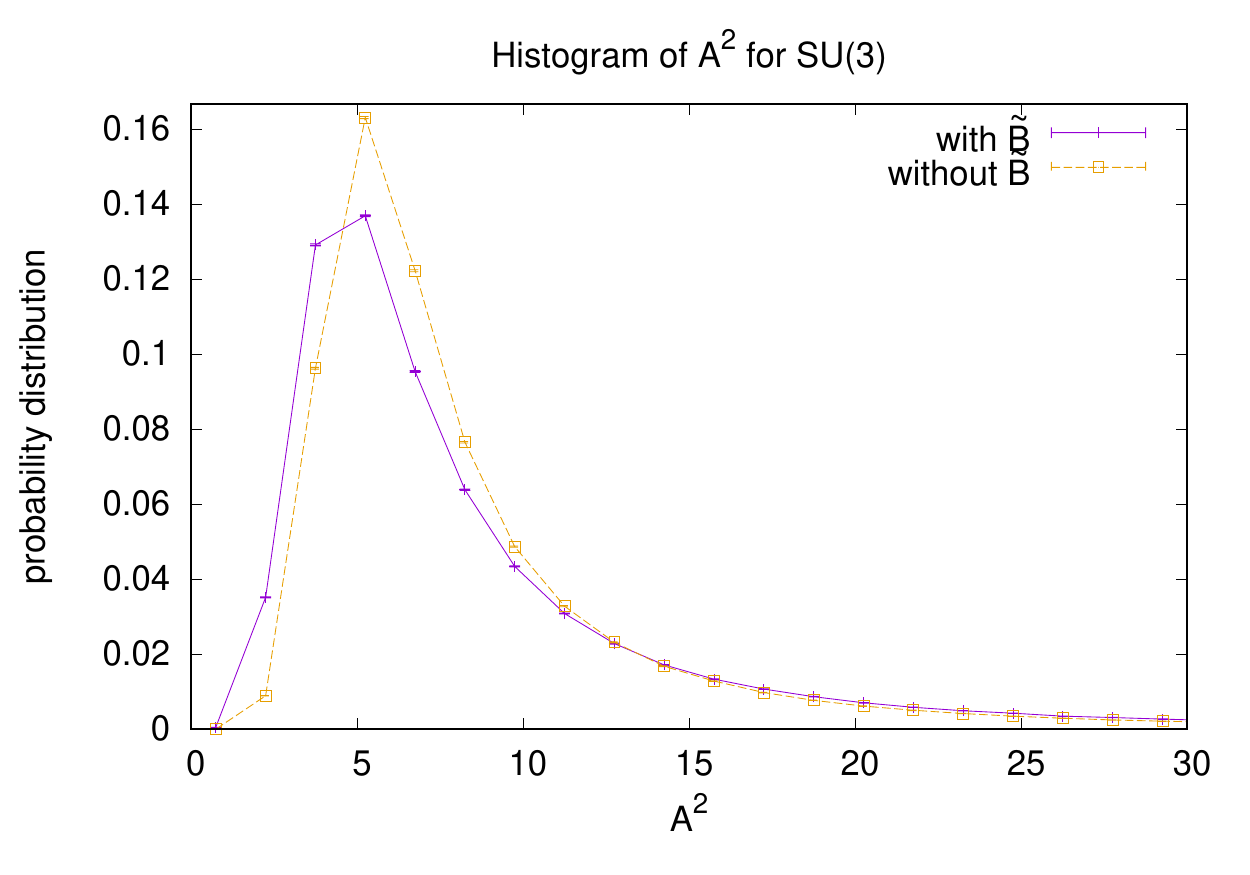}
 \includegraphics[scale=0.62]{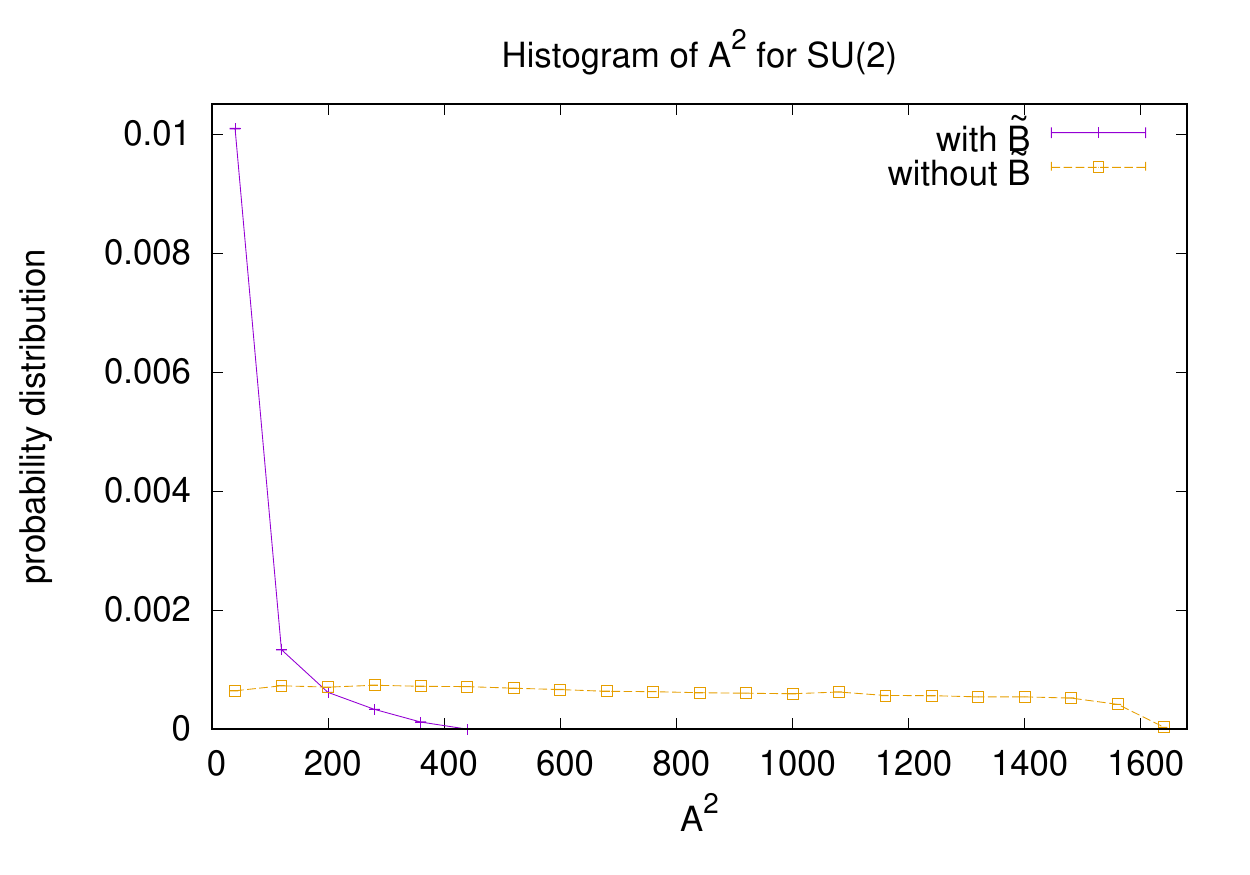}
 \caption{Histograms of $\frac{1}{N}\tr (A_\mu)^2$ 
obtained by simulating 
the zero-mode effective theory for $N=3$ (Left) and $N=2$ (Right).
We also plot the results obtained by 
setting the fluctuations $\tilde B_\mu$ in the toron to zero for comparison.
 }
 \label{fig:Asq_for_zero-mode}
\end{figure}

In Fig.~\ref{fig:Asq_for_zero-mode}
we plot
the 
histogram
of $\tr (A_\mu)^2$ for $N=2$ and $N=3$,
which shows that
the results
for $N=2$
has a much longer tail than that 
for $N=3$.
In the same figure, we also plot the 
histogram
obtained
by setting $\tilde B_\mu$ in $X_\mu$ to zero,
which corresponds to prohibiting
the fluctuations in the toron.
We observe
that
the 
histogram
in the SU(2) case
becomes almost flat
with a sudden plunge at $\frac{1}{N}\tr (A_\mu)^2\sim 1600$,
which seems to indicate some problem in the simulation in this case.
In any case, it is
conceivable that
the large fluctuations in the zero mode $A_\mu$ 
that appear around the trivial vacuum $\tilde B_\mu = 0$
are responsible for the dominance 
of 
the trivial vacuum $\tilde B_\mu = 0$.

\begin{figure}[htbp]
 \centering
 \includegraphics[scale=0.62]{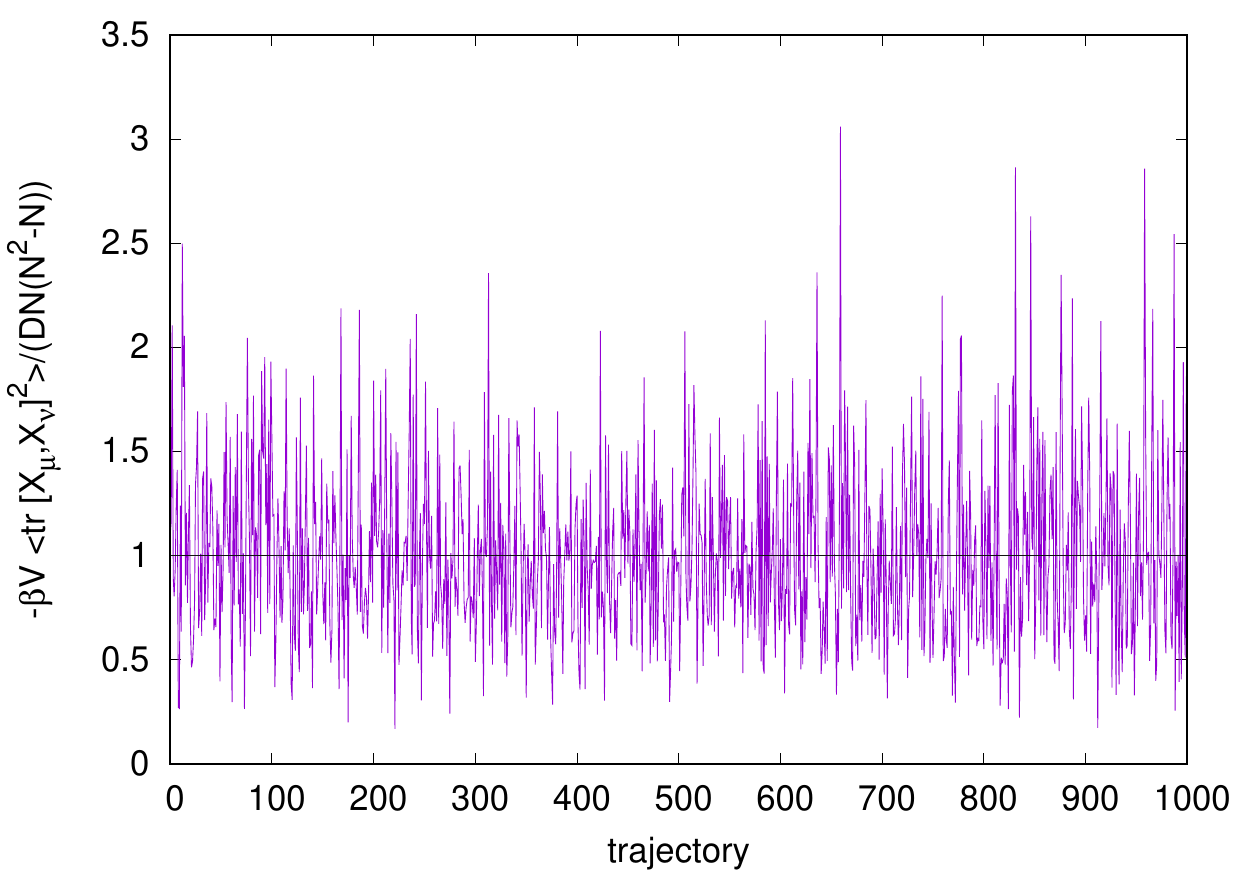}
 \includegraphics[scale=0.62]{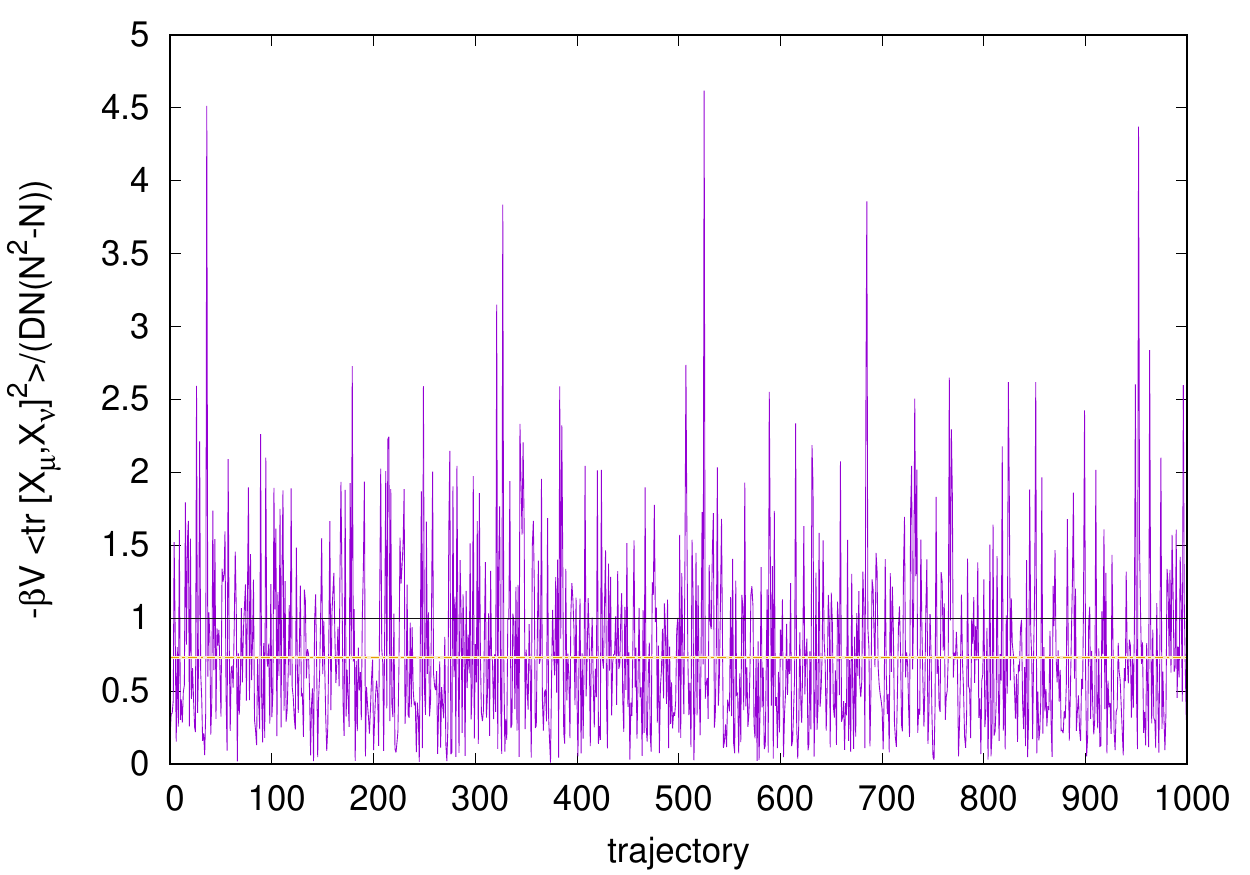}
 \caption{Histories of the quantity that is supposed to be unity theoretically
due to the Schwinger-Dyson equation
in the zero-mode effective theory for $N=3$ (Left) and $N=2$ (Right).
The black horizontal line indicates the exact value,
whereas the dashed line with a band indicates the measured 
expectation value with a statistical error.
The two types of line coincide in the $N=3$ case.}
 \label{fig:virial_for_zero-mode}
\end{figure}

The problem in the simulation for $N=2$ can be seen also by checking
the Schwinger-Dyson equation \eqref{SDeq_for_zero-mode}.
In the case of setting $\tilde B_\mu=0$,
the right-hand side of Eq.~\eqref{SDeq_for_zero-mode}
should be replaced by $D(N^2-N)$.
In table~\ref{tab:virial} we show the left-hand side of
the Schwinger-Dyson equation normalized by the constant 
on the right-hand side so that 
one obtains unity if the Schwinger-Dyson equation is satisfied.
We find that the result is indeed unity with the error bar for $N=3$,
but not for $N=2$.
In Fig.~\ref{fig:virial_for_zero-mode}
we plot the history of this quantity
in the case without the fluctuations $\tilde B_\mu$ in the toron,
which corresponds to the case with larger deviation from unity
for the $N=2$ case as one can see from table~\ref{tab:virial}.
We observe that the history for $N=2$ has larger spikes than that for $N=3$.
From this observation with the fact that the expectation value
is smaller than theoretically expected, it is conceivable
that
the simulation fails to sample rare configurations that give large spikes.
We consider that this is due to the large fluctuations in $A_\mu$
that appear when the fluctuations $\tilde B_\mu$ in the toron become small.

\begin{table}[htbp]
 \centering
 \caption{The expectation values of $-\frac{\beta V}{DN}\sum_{\mu \ne \nu}\langle\tr[X_\mu, X_\nu]^2\rangle$ normalized by $N^2-1$ or by $N^2-N$ 
depending on whether $\tilde B_\mu$ is included or not.}
 \begin{tabular}{c|c|c}
  \hline\hline
  &$N=3$&$N=2$\\
  \hline
  w/o $\tilde B_\mu$&0.999(3) &0.729(8) \\
  w/ $\tilde B_\mu$&1.0009(12) &0.966(4) \\
  \hline\hline
 \end{tabular}
 \label{tab:virial}
\end{table}%

These observations can be understood by 
recalling the dynamical properties of
the reduced bosonic model \eqref{reducedYM-action}.
In particular, it was found that 
%
%
the eigenvalue distribution $\rho(x)$ of one of the $D$ matrices, say $X_1$,
in the model, has a power-law tail \cite{KS99Ei}
\begin{align}
  \rho(x)\sim x^{-p} \ , \quad \quad  p=2N(D-2)-3D+5 
\label{power-law-EV}
\end{align}
except for $D=N=3$.
Note that, in $D=4$, the power is $p=1$ for $N=2$ and $p=5$ for $N=3$.
Since the partition function is given by
$Z \sim \int_{-\infty}^{\infty} dx \, \rho(x)$,
it converges for $N=3$, but diverges logarithmically\footnote{The
logarithmic divergence of the partition function \eqref{reducedYM-action}
for $N=2$ was found earlier in Ref.~\cite{Coste:1985mn} from a different
argument. However, the dominance of the trivial vacuum was not concluded.}
for $N=2$.
When $B_\mu \neq 0$, one obtains
quadratic terms in $A_\mu$ from \eqref{plaquette-expansion},
which suppress the fluctuations in $A_\mu$.
Therefore, in the $N=2$ case, the huge fluctuations in $A_\mu$ appear
only when $B_\mu=0$ and make 
$B_\mu=0$ dominate at large $\beta$.

Next we discuss the slow convergence to the trivial vacuum,
which is found to be described by the $1/\log \beta$ corrections
in Fig.~\ref{fig:b1-WL_Nc2_Ls4_Lt4}.
Note first that,
the fluctuations in $A_\mu$ is bounded
by O($\beta^{-1/4}$) 
according to the effective theory~\eqref{Sg-expansion}
at large $\beta$ even when $B_\mu=0$.
Therefore, the fluctuations in $A_\mu$ cannot be the reason for
the slow convergence, and hence we focus on the fluctuations in $B_\mu$,
which will be denoted in what follows as $B$ for simplicity.
Since the quadratic term in $(A_\mu)_{12}$ in the action has 
a coefficient $((B_\mu)_{11}-(B_\mu)_{22})^2\sim B^2$,
the partition function for $B_\mu$ after integrating out $A_\mu$
is suppressed at large $B$ by $1/B^2$.
At small $B$, on the other hand,
the partition function for $B_\mu$
is expected to have a bound $\int dx\,\rho(x)\sim \ln\beta$
due to the power-law behavior \eqref{power-law-EV}
since $\beta$ plays the role of a cutoff
for the zero-mode effective theory.
Combining these two asymptotic behaviors of 
the partition function for $B_\mu$,
the typical value of $B$ 
can be roughly estimated by
balancing
$1/B^2$ and $\ln\beta$.
This explains the slow convergence to the trivial vacuum.
\section{Polyakov line v.s.\ perturbative predictions}
\label{sec:polyakov}

In this section
we provide further supports on our conclusion
by measuring the Polyakov line for SU(2) and SU(3)
on the $4^4$ lattice with the same setup as in section \ref{sec:wilson-loop}.
In particular, we show that 
the probability distribution of the Polyakov line
exhibits a remarkable difference,
which can be understood by the discussion in the previous section.

\begin{figure}[htbp]
 \centering
  \includegraphics[scale=0.6]{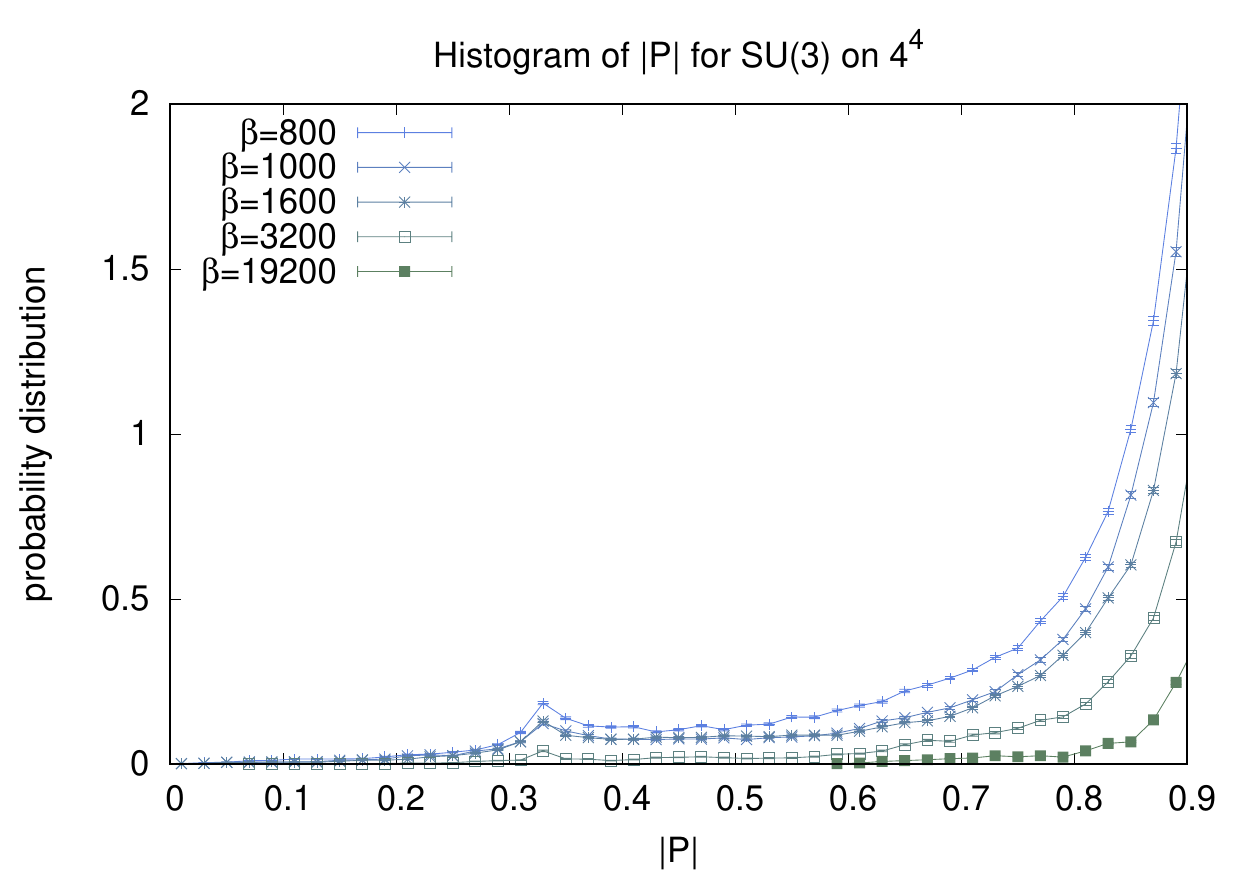}
 \includegraphics[scale=0.6]{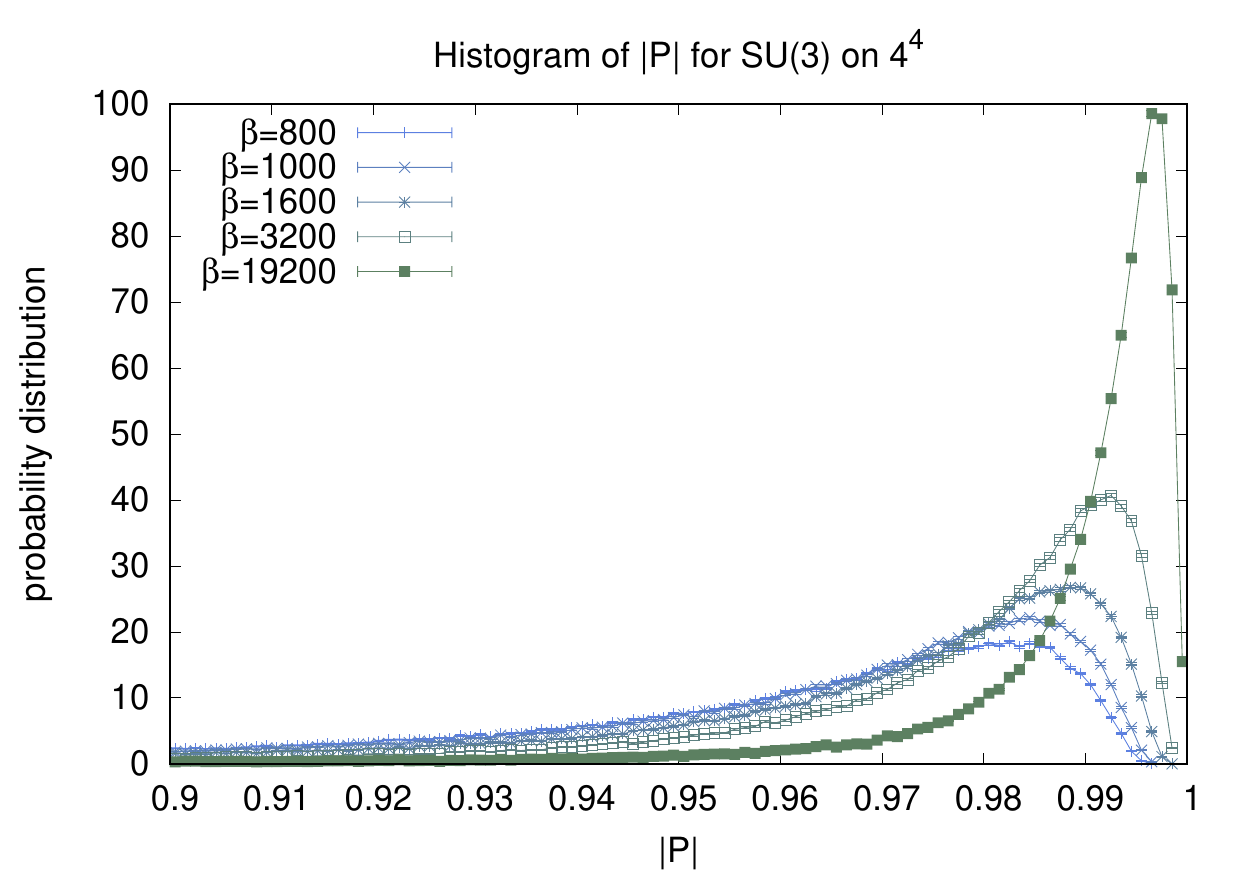}
  \includegraphics[scale=0.6]{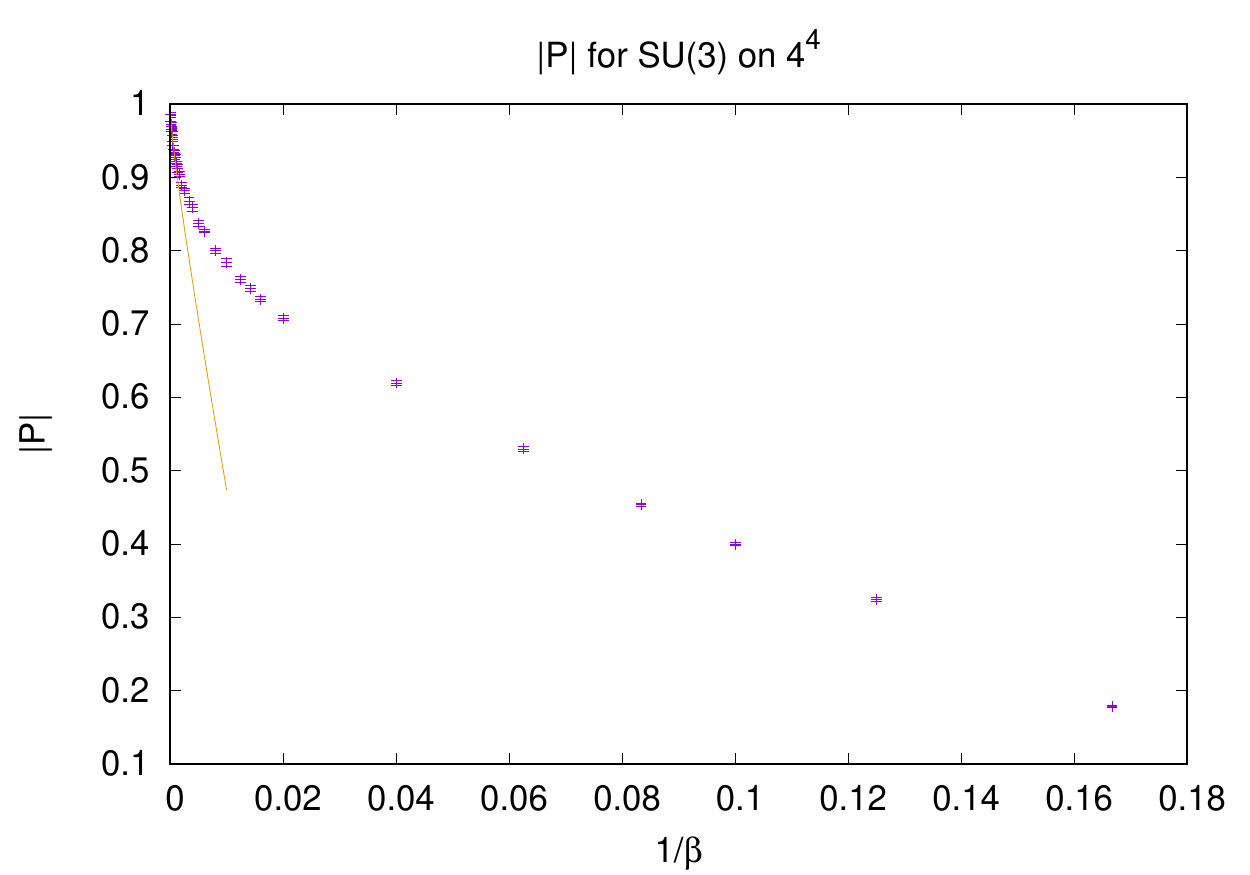}
 \includegraphics[scale=0.6]{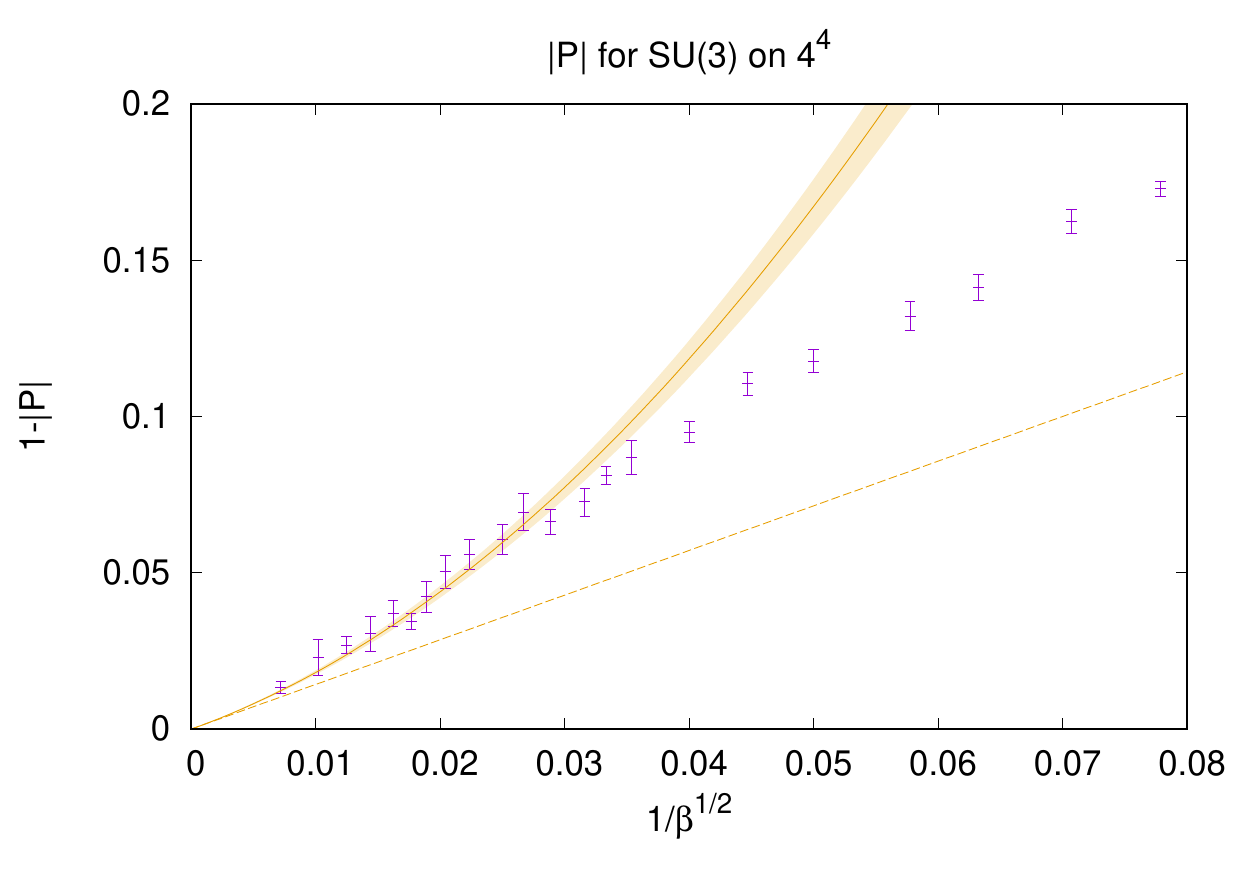}
 \caption{(Top) Histograms of $|P|$ for $N=3$ on the $4^4$ lattice. 
In the left panel, we show the region $0 \le |P| \le 0.9$
with the bin size 0.02,
while in the right panel, we show the region $0.9 \le |P| \le 1$
with the bin size 0.001. 
(Bottom-Left) $|P|$ for $N=3$ on the $4^4$ lattice plotted against
   $1/\beta$. (Bottom-Right)
   $1-|P|$ for $N=3$ on the $4^4$ lattice
 plotted against $1/\sqrt{\beta}$.
The dashed line represents the leading perturbative
prediction \eqref{Pol-prediction}.
The solid line represents a fit assuming O($\beta^{-1}$) corrections.
 }
 \label{fig:hist_absP_Nc3_Ls4_Lt4}
\end{figure}

\begin{figure}[htbp]
  \centering
  \includegraphics[scale=0.6]{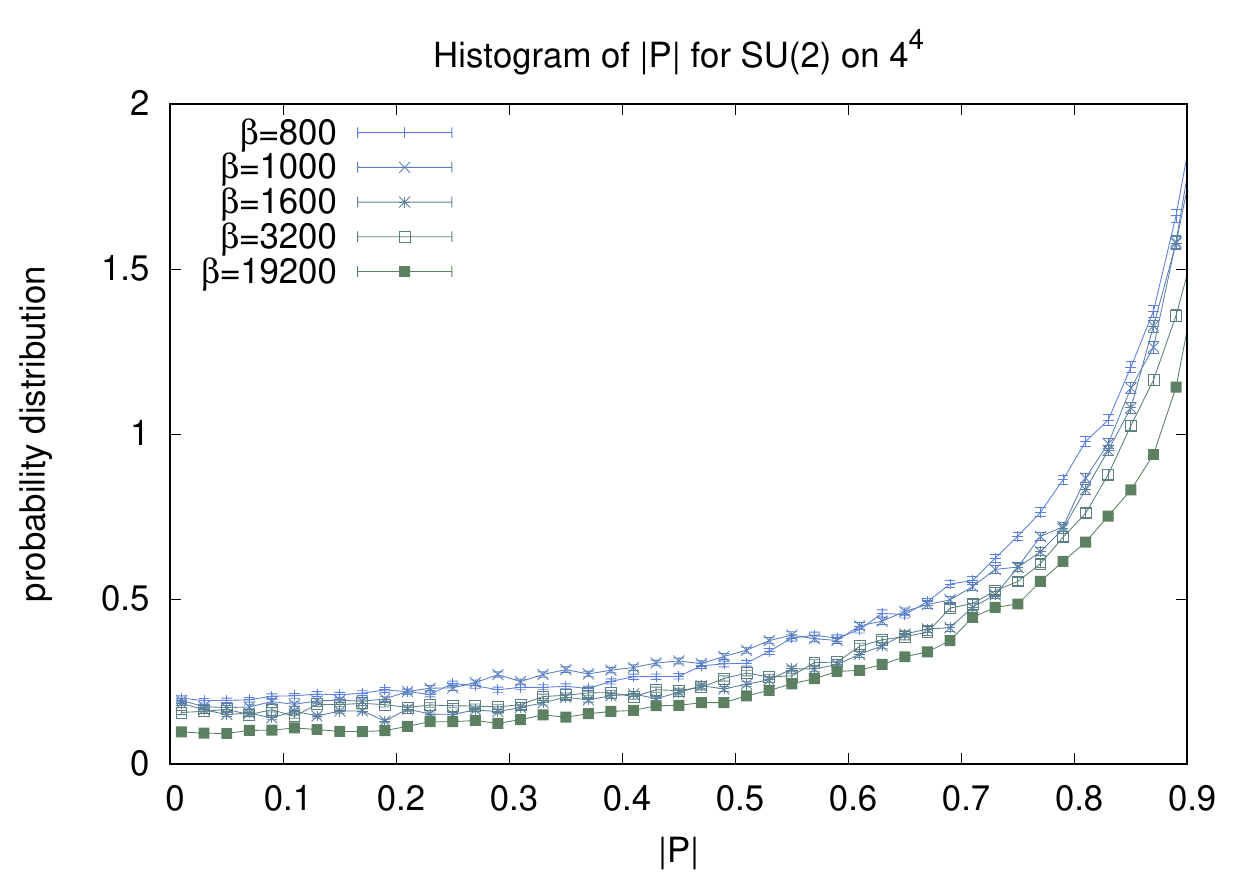}
 \includegraphics[scale=0.6]{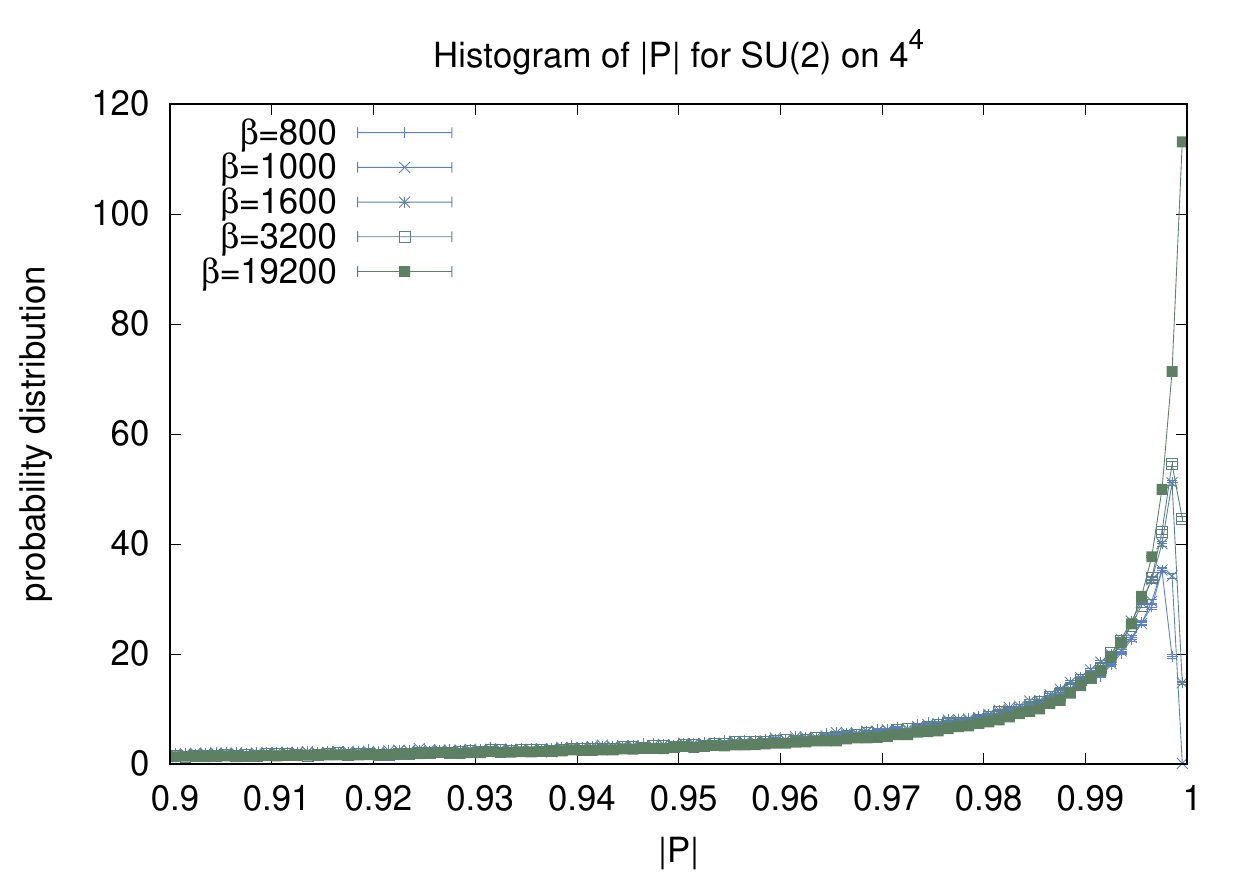}
 \includegraphics[scale=0.6]{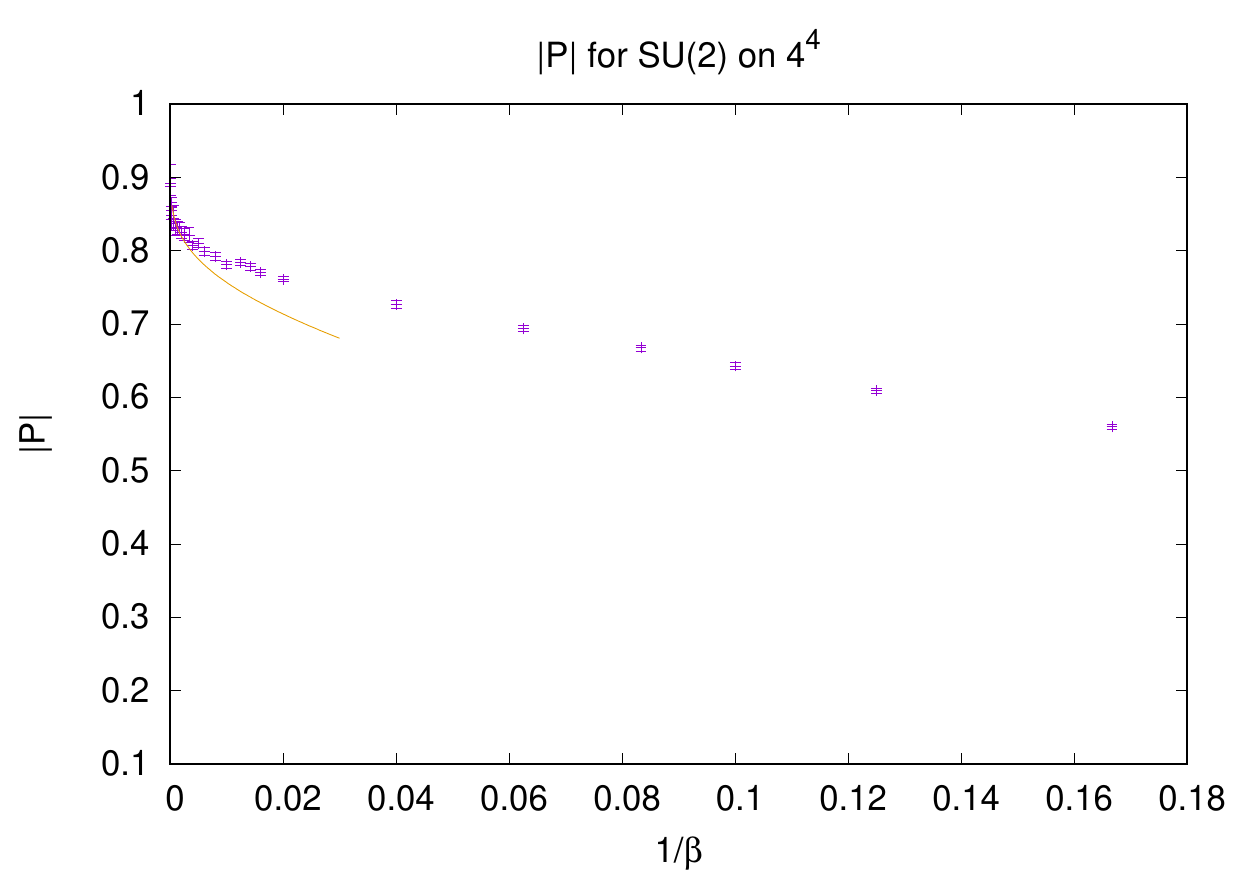}
 \includegraphics[scale=0.6]{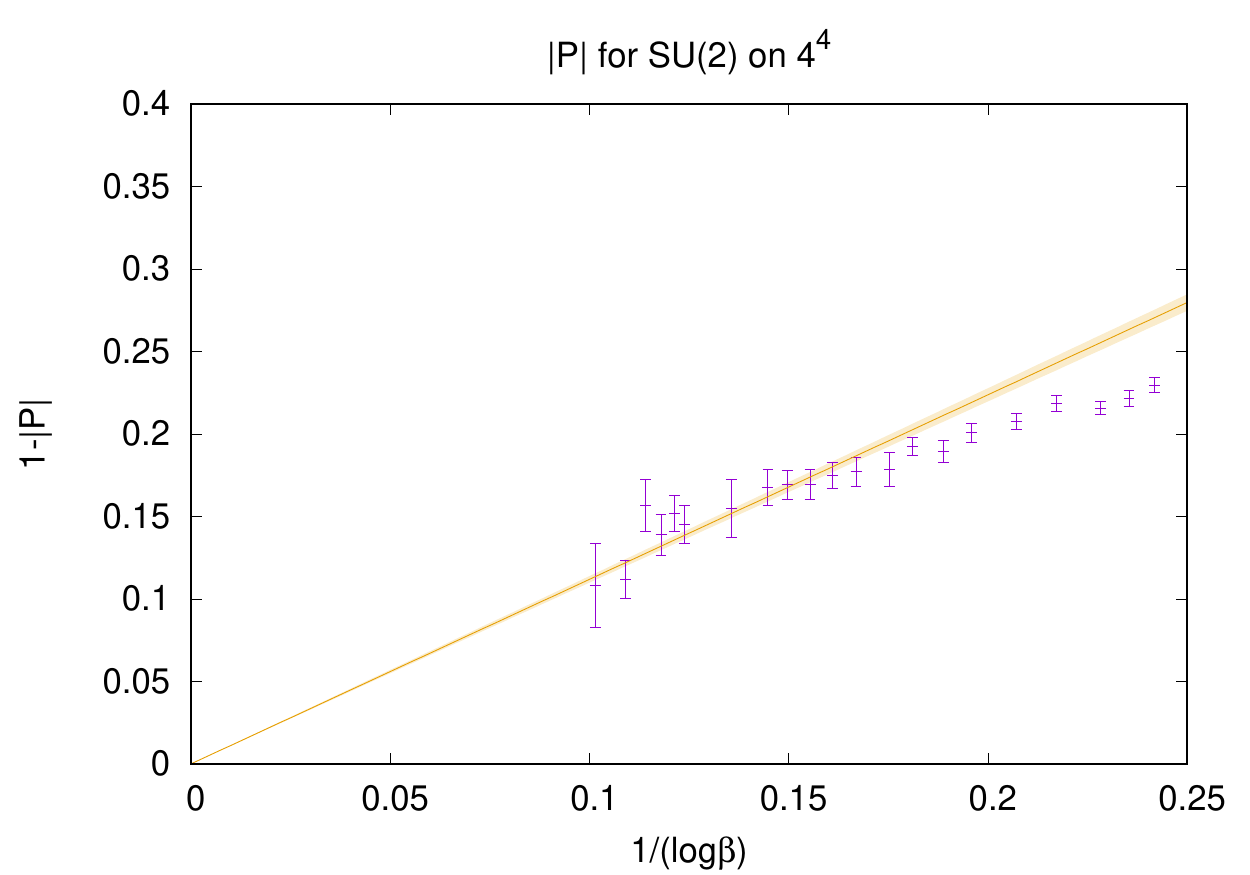}
 \caption{(Top) Histograms of $|P|$ for $N=2$ on the $4^4$ lattice. 
In the left panel, we show the region $0 \le |P| \le 0.9$
with the bin size 0.02,
while in the right panel, we show the region $0.9 \le |P| \le 1$
with the bin size 0.001. 
(Bottom-Left) $|P|$ for $N=2$ on the $4^4$ lattice plotted against
   $1/\beta$. (Bottom-Right)
   $1-|P|$ for $N=2$ on the $4^4$ lattice
 plotted against $1/\log \beta$.
The solid line represents a fit assuming O($1/\log \beta$) corrections.
}
 \label{fig:hist_absP_Nc2_Ls4_Lt4}
\end{figure}

In Figure \ref{fig:hist_absP_Nc3_Ls4_Lt4}, we present our results
for SU(3) on the $4^4$ lattice.
In the Top panels, we show
the histogram of the Polyakov line $|P|$ for various $\beta$
within the region $|P|\le 0.9$ (Left) and $|P| \ge 0.9$ (Right), 
respectively.
The peak near
$|P|=1$ becomes sharper and sharper for increasing $\beta$,
which confirms that the trivial vacuum dominates and the zero-mode
fluctuations are suppressed in the large $\beta$ limit.


\mycomment{
For sufficiently large $\beta$, 
the Polyakov line can be approximated
by the trace of zero modes as $P=\frac{1}{N}\tr U_{4}^{L_4}$.
Denoting the eigenvalues of $U_{4}$ by $e^{i\theta_i}$,
$|P|$ can be expressed as
\begin{align}
 |P|\simeq \left| \frac{1}{N}\sum_{i=1}^{N}e^{iL_4\theta_i} \right|
 =1-\frac{L_4{}^2}{2N}\sum_{i=1}^{N}\theta_i{}^2+ \cdots
 \  .
\end{align}
Thus $\sqrt{1-|P|}$ represents
the standard deviation of the distribution of $\theta_i$,
which is a good measure of the zero-mode fluctuations.
In the Top-Right panel, we show the histogram of the $\sqrt{1-|P|}$.
We find 
a peak near the origin,
which becomes sharper and
comes closer to the origin with increasing $\beta$.
In the Bottom-Left panel,
we plot the expectation value of $|P|$,
which approaches one with increasing $\beta$.
}

Using the perturbative expansion
around $B_\mu=0$,
we obtain a prediction for the Polyakov line as
\begin{align}
  P&= 1-\frac{L_4{}^2}{2N}
  \left\langle \tr\left[ (A_4+\tilde{B}_4)^2 \right]\right\rangle
  +\frac{L_4{}^4}{4! \, N}\left\langle \tr\left[ (A_4+\tilde{B}_4)^4
    \right]\right\rangle
 \nonumber \\
&  \quad
 -\frac{N^2-1}{2\beta V}\sum_{k\ne 0}\frac{1}{4\sum_\mu \sin^2\frac{k_\mu}{2}}
 \left| \sum_{m=0}^{L_4-1} e^{ik_4m} \right|^2
 +{\rm O}(\beta^{-\frac{3}{2}}) \  .
\label{Pol-prediction}
\end{align}
The last term in the first line is logarithmically 
divergent for $N=3$ due to the
power-law tail \eqref{power-law-EV}.
In the full lattice theory, this divergence is replaced
by unpredictable O($\beta^{-1}\log\beta$) terms,
which are indistinguishable from the O($\beta^{-1}$) corrections.
In the Bottom-Right panel, we plot $1-|P|$ against $1/\sqrt{\beta}$
and compare our results against the O($\beta^{-1/2}$)
perturbative prediction.
We find that our results can be fitted 
by assuming O($\beta^{-1}$) corrections
as expected from the above argument.





In Fig.~\ref{fig:hist_absP_Nc2_Ls4_Lt4},
we show our results for the $N=2$ case.
From the Top panels, we find that
the histogram of $|P|$ has
a peak near $|P|=1$, which becomes sharper and sharper for increasing $\beta$.
However, the difference from the SU(3) case lies in the tail of the
histogram, where the height of the histogram
is decreasing but only slowly.
From the Bottom-Left panel,
it is not even clear whether 
the expectation value of $|P|$
is approaching unity as $\beta$ increases.
From the Bottom-Right panel,
we find that $1-|P|$ decreases as $1/\log{\beta}$ at large $\beta$,
which is similar to the behavior of the Wilson loop
in Fig.~\ref{fig:b1-WL_Nc2_Ls4_Lt4} (Right).
These behaviors are consistent with 
our conclusion in section \ref{sec:reduced-YM}
that
the trivial vacuum dominates at large $\beta$,
while the fluctuations in the zero modes $A_\mu$ are large
and decrease very slowly with increasing $\beta$.





\section{Summary and discussions}
\label{sec:summary}

In this paper 
we discussed a subtle issue in lattice perturbation theory
for 4D SU(2) and SU(3) gauge theories with periodic boundary conditions.
The minima of the action are actually degenerate and they are given
by the toron configurations.
For the SU(3) case, it is known that the trivial vacuum 
dominates over the nontrivial torons as $\beta \rightarrow \infty$,
and the fluctuations of the zero modes are described by the effective theory,
which takes the form of the bosonic reduced model.
In the SU(2) case, the criterion on the dominance of the trivial vacuum 
discussed in the previous work is marginal.
We have found in this case that
the Wilson loops at large $\beta$ reproduce the 
leading $1/\beta$ correction obtained perturbatively around the trivial 
vacuum as shown in Fig.~\ref{fig:b1-WL_Nc2_Ls4_Lt4}
although the subleading term decreases
very slowly except for the $1 \times 1$ Wilson loop.

We have provided an explanation of this observation
based on the zero-mode effective theory.
Namely, the trivial vacuum dominates because the fluctuations in the zero modes
become large around it as one can deduce from the power-law
tail of the eigenvalue distribution 
in the reduced model.
We have confirmed our conclusion by 
measuring the Polyakov line, which shows the dominance of the trivial vacuum
as well as the large fluctuations in the zero modes at large $\beta$. 



The issue addressed in this paper
concerns any SU($N$) gauge theory in a finite periodic box.
Note, however, that it becomes irrelevant in the infinite-volume limit
in which the boundary conditions have no effects.
This can be understood 
since
the gauge symmetry and the center symmetry 
allows us to restrict
the integration domain of $B_\mu$
to \cite{Coste:1985mn}\footnote{In fact,
one can use the center symmetry 
to restrict the
integration domain of $B_\mu$ to
\begin{align}
  -\frac{N-1}{N}\frac{\pi}{L_\mu} < (B_\mu)_{ii} \le
  \frac{N-1}{N}\frac{\pi}{L_\mu}    \ ,
\end{align}
as one can prove explicitly for $N=2$ and $3$.
However, this does not affect our discussion here.
}
\begin{align}
-\frac{2\pi}{L_\mu}< (B_\mu)_{ii} - (B_\mu)_{jj} < \frac{2\pi}{L_\mu} \ ,
\end{align}
which shrinks to zero
in the $L_\mu \rightarrow \infty$ limit.
On the other hand,
the issue is relevant in a finite periodic box\footnote{Such a set up
is useful in making perturbative analyses applicable.
For instance, 
QCD in a small box 
has been used recently in obtaining
the critical point for color superconductivity based on
the one-loop self-consistency equation \cite{Yokota:2023osv}.} 
\emph{even in the continuum limit},
in which
the toron configurations should be described
by $B_\mu^{\text{cont}} \equiv B_\mu/a$, where $a$ is the lattice spacing,
since the integration domain of $B_\mu^{\text{cont}}$ remains finite
in the $L_\mu \rightarrow \infty$ limit with $L_\mu a$ fixed.

The subtlety in the 4D SU(2) gauge theory
discussed in this paper
is expected to be important also
when one performs Monte Carlo simulations in a small box
since the simulations may suffer from the ergodicity problem
associated with the nontrivial dynamics of the zero modes.
For instance, it was reported that 
the ``two-color QCD'' simulations
on a $3^3 \times 64$ lattice with $N_{\rm f}=2$ flavor Wilson quarks
at finite density failed to reproduce the behavior of the
quark number density expected for free quarks \cite{Hands:2010vw}.
We suspect that there might be some problem in the simulations
judging from the fact that our independent simulations 
with the same setup reproduce the expected behavior.

Finally let us recall that
the trivial vacuum is expected \emph{not} to dominate
for the 3D SU(2) gauge theory
according to the 
criterion 
of Ref.~\cite{Coste:1985mn} (See footnote \ref{footnote:3d-SU2}.).
Note, however, that
the partition function 
of the naive zero-mode effective theory \eqref{reducedYM-action}
diverges 
according to the eigenvalue distribution \eqref{power-law-EV}
for $D=3$ and $N=2$.
This effect, which clearly favors the trivial vacuum, 
is not taken
into account in the criterion of Ref.~\cite{Coste:1985mn}.
Therefore we think it worth while to simulate this theory
in a small box to see whether the prediction \eqref{WL-expansion}
around the trivial vacuum is approached or not in the large $\beta$ limit.



\section*{Acknowledgment}
We would like to thank Etsuko Itou for providing us with some
results in 4D SU(2) gauge theory at finite density.
We are also grateful to Yuta Ito, Hideo Matsufuru, Yusuke Namekawa,
Asato Tsuchiya, Shoichiro Tsutsui and Takeru Yokota
for discussions on finite density QCD,
which motivated the present work.
The computations were carried out on
the PC clusters in KEK Computing Research Center
and KEK Theory Center.
We have used Bridge++ (http://bridge.kek.jp/Lattice-code/), which is a code set for numerical simulations of 
lattice gauge theories based on C++ \cite{Ueda:2014rya,Akahoshi:2021gvk}.
We would like to thank Hideo Matsufuru for his help concerning the usage
of this code set.

\bibliographystyle{JHEP}
\bibliography{0-mode}

\providecommand{\href}[2]{#2}\begingroup\raggedright\begin{thebibliography}{10}

\bibitem{Eguchi:1982nm}
T.~Eguchi and H.~Kawai, \emph{{Reduction of Dynamical Degrees of Freedom in the
  Large N Gauge Theory}},
  \href{https://doi.org/10.1103/PhysRevLett.48.1063}{\emph{Phys. Rev. Lett.}
  {\bfseries 48} (1982) 1063}.

\bibitem{Gonzalez-Arroyo:1982hyq}
A.~Gonzalez-Arroyo and M.~Okawa, \emph{{The Twisted Eguchi-Kawai Model: A
  Reduced Model for Large N Lattice Gauge Theory}},
  \href{https://doi.org/10.1103/PhysRevD.27.2397}{\emph{Phys. Rev. D}
  {\bfseries 27} (1983) 2397}.

\bibitem{Gonzalez-Arroyo:2010omx}
A.~Gonzalez-Arroyo and M.~Okawa, \emph{{Large $N$ reduction with the Twisted
  Eguchi-Kawai model}},
  \href{https://doi.org/10.1007/JHEP07(2010)043}{\emph{JHEP} {\bfseries 07}
  (2010) 043} [\href{https://arxiv.org/abs/1005.1981}{{\ttfamily 1005.1981}}].

\bibitem{Ishibashi:1996xs}
N.~Ishibashi, H.~Kawai, Y.~Kitazawa and A.~Tsuchiya, \emph{{A Large N reduced
  model as superstring}},
  \href{https://doi.org/10.1016/S0550-3213(97)00290-3}{\emph{Nucl. Phys. B}
  {\bfseries 498} (1997) 467}
  [\href{https://arxiv.org/abs/hep-th/9612115}{{\ttfamily hep-th/9612115}}].

\bibitem{Anagnostopoulos:2022dak}
K.~N. Anagnostopoulos, T.~Azuma, K.~Hatakeyama, M.~Hirasawa, Y.~Ito,
  J.~Nishimura et~al., \emph{{Progress in the numerical studies of the type IIB
  matrix model}},  10, 2022, \href{https://arxiv.org/abs/2210.17537}{{\ttfamily
  2210.17537}}.

\bibitem{Coste:1985mn}
A.~Coste, A.~Gonzalez-Arroyo, J.~Jurkiewicz and C.~P. Korthals~Altes,
  \emph{{Zero Momentum Contribution to Wilson Loops in Periodic Boxes}},
  \href{https://doi.org/10.1016/0550-3213(85)90064-1}{\emph{Nucl. Phys.}
  {\bfseries B262} (1985) 67}.

\bibitem{Fodor:2012td}
Z.~Fodor, K.~Holland, J.~Kuti, D.~Nogradi and C.~H. Wong, \emph{{The Yang-Mills
  gradient flow in finite volume}},
  \href{https://doi.org/10.1007/JHEP11(2012)007}{\emph{JHEP} {\bfseries 11}
  (2012) 007} [\href{https://arxiv.org/abs/1208.1051}{{\ttfamily 1208.1051}}].

\bibitem{Fodor:2012qh}
Z.~Fodor, K.~Holland, J.~Kuti, D.~Nogradi and C.~H. Wong, \emph{{The gradient
  flow running coupling scheme}},
  \href{https://doi.org/10.22323/1.164.0050}{\emph{PoS} {\bfseries LATTICE2012}
  (2012) 050} [\href{https://arxiv.org/abs/1211.3247}{{\ttfamily 1211.3247}}].

\bibitem{KS99Ei}
W.~Krauth and M.~Staudacher, \emph{Eigenvalue distributions in yang-mills
  integrals},
  \href{https://doi.org/10.1016/S0370-2693(99)00395-0}{\emph{Phys.Lett.}
  {\bfseries B453} (1999) 253}
  [\href{https://arxiv.org/abs/hep-th/9902113}{{\ttfamily hep-th/9902113}}].

\bibitem{Yokota:2023osv}
T.~Yokota, Y.~Ito, H.~Matsufuru, Y.~Namekawa, J.~Nishimura, A.~Tsuchiya et~al.,
  \emph{{Color superconductivity on the lattice -- analytic predictions from
  QCD in a small box}},  \href{https://arxiv.org/abs/2302.11273}{{\ttfamily
  2302.11273}}.

\bibitem{Hands:2010vw}
S.~Hands, T.~J. Hollowood and J.~C. Myers, \emph{{Numerical Study of the Two
  Color Attoworld}}, \href{https://doi.org/10.1007/JHEP12(2010)057}{\emph{JHEP}
  {\bfseries 12} (2010) 057} [\href{https://arxiv.org/abs/1010.0790}{{\ttfamily
  1010.0790}}].

\bibitem{Ueda:2014rya}
S.~Ueda, S.~Aoki, T.~Aoyama, K.~Kanaya, H.~Matsufuru, S.~Motoki et~al.,
  \emph{{Development of an object oriented lattice QCD code 'Bridge++'}},
  \href{https://doi.org/10.1088/1742-6596/523/1/012046}{\emph{J. Phys. Conf.
  Ser.} {\bfseries 523} (2014) 012046}.

\bibitem{Akahoshi:2021gvk}
Y.~Akahoshi, S.~Aoki, T.~Aoyama, I.~Kanamori, K.~Kanaya, H.~Matsufuru et~al.,
  \emph{{General purpose lattice QCD code set Bridge++ 2.0 for high performance
  computing}}, \href{https://doi.org/10.1088/1742-6596/2207/1/012053}{\emph{J.
  Phys. Conf. Ser.} {\bfseries 2207} (2022) 012053}
  [\href{https://arxiv.org/abs/2111.04457}{{\ttfamily 2111.04457}}].

\end{thebibliography}\endgroup

\end{document}